\documentclass{emulateapj}

\def\amin{\ifmmode^{\prime}\else$^{\prime}$\fi}
\def\asec{\ifmmode^{\prime\prime}\else$^{\prime\prime}$\fi}

\shorttitle{AGN in Nuclear Star Clusters}
\shortauthors{}

\begin{document}

\slugcomment{Accepted by ApJ, Dec. 20, 2007}

\title{The Coincidence of Nuclear Star Clusters and Active Galactic Nuclei}

\author{Anil Seth\footnote{CfA Fellow}}
\affil{Harvard-Smithsonian Center for Astrophysics}
\email{aseth@cfa.harvard.edu}

\author{Marcel Ag\"ueros\footnote{NSF Astronomy and Astrophysics Postdoctoral Fellow}}
\affil{Columbia University}

\author{Duane Lee}
\affil{Columbia University}

\author{Antara Basu-Zych}
\affil{Columbia University}

\begin{abstract}

We study galaxies that host both nuclear star clusters and active
galactic nuclei (AGN) implying the presence of a massive black hole.
We select a sample of 176 galaxies with previously detected nuclear
star clusters that range from ellipticals to
late-type spirals.  We search for AGN in this sample using
optical spectroscopy and archival radio and X-ray data.  We find 
galaxies of all Hubble types and with a
wide range of masses ($10^{9}-10^{11}$~M$_\odot$) hosting both AGN and 
nuclear star clusters.  From the optical
spectra, we classify 10\% of the galaxies as AGN and an additional
15\% as composite, indicating a mix of AGN and star-formation spectra.
The fraction of nucleated galaxies with AGN increases strongly as a
function of galaxy and nuclear star cluster mass.  For galaxies with
both a NC and a black hole, we find that the masses of these two objects
are quite similar.  However, non-detections of black holes in Local
Group nuclear star clusters show that not all clusters host black
holes of similar masses.  We discuss the implications of our results for
the formation of nuclear star clusters and massive black holes.

\end{abstract}


\section{Introduction}

Nuclear star clusters (NCs) are massive star clusters coincident with
the photocenters of galaxies.
They are very common and have been found in $\sim$75\% of local
late-type spirals \citep{boker02} and Virgo dwarf elliptical galaxies
\citep{cote06}.  Their size is similar to that of globular clusters
\citep{boker04a}, but NCs are 1-2 orders of magnitude brighter and
more massive \citep{walcher05}.  Also, unlike most globular clusters,
they have extended star formation histories \citep{walcher06,rossa06}
and complex morphologies \citep{seth06}.

The luminosity of nuclear star clusters correlates with
galaxy luminosity in both ellipticals \citep{lotz01,graham03,cote06}
and spirals \citep{carollo98,boker04a}.  Recently, it has been
shown that the masses of NCs follow scaling relationships with galaxy
mass ($M_{gal}$), bulge velocity dispersion ($\sigma$), and S\'ersic
index \citep{ferrarese06,wehner06,rossa06,graham07}.  These scaling
relations are very similar to those seen for massive black holes
(MBHs\footnote{We use massive black holes to refer to all
non-stellar-mass black holes, including those normally referred to as
intermediate mass and as super-massive black holes.}), appearing to
extend those relations to lower masses.  Thus,
\citet{wehner06} and \citet{ferrarese06} suggest that there may be a
single scaling relation linking the mass of a central massive object
(CMO; either a NC or MBH) to the large-scale properties of the galaxy.
The existence of an $M_{CMO}-\sigma$ or $M_{CMO}-M_{gal}$ relation
suggests that the formation of the CMO is linked in some way to the
evolution of the galaxy.  Theoretically, these scaling relationships
can be understood in multiple ways.  They could be created by feedback
from either NCs or MBHs regulating the star formation in the galaxy
as a whole \citep[e.g.,][]{mclaughlin06b}, or alternatively, they could
simply result from gas accretion onto the nucleus in proportion to the
galaxies' mass \citep[e.g.,][]{li07}.

Despite this interesting connection between NCs and MBHs, and their
link to galaxy formation and evolution, no systematic study of the 
overlap between these classes of objects exists.
A handful of
objects, including the Milky Way, are already known to host both NCs
and MBHs (see \S\ref{litsec}).  \citet{ferrarese06} and
\citet{wehner06} show there is a rough transition at galaxy masses of
$\sim$10$^{10}$~M$_\odot$ (and corresponding CMO mass of
$\sim$10$^7$~M$_\odot$), above which galaxies typically host MBHs, and
below which galaxies have NCs.  While there is good evidence that more
massive galaxies do not in fact host NCs \citep{cote06}, it remains
unclear how common MBHs are in lower mass galaxies
\citep[e.g.,][]{greene07}.  Recent theoretical work shows that MBHs
could form from stellar mergers in a young, dense cluster environment
\citep{miller04,portegieszwart04}, and a direct link between NC and
MBH formation may therefore exist.

We present a systematic study of the overlap between NCs and MBHs
aimed at better understanding the relation between the two types of
objects and the formation mechanism of CMOs in general.
Starting with a sample of galaxies with known NCs, we search for
active galactic nuclei (AGN) that are powered by accretion onto an MBH.
This study gives a lower limit on the number of systems with MBHs,
since quiescent and heavily obscured MBHs will not be detected as AGN.

We begin by describing our sample of galaxies with NCs, drawn from
several different catalogs (\S\ref{samplesec}). Using optical
spectra and radio and X-ray data, we examine our sample galaxies for
evidence of AGN activity (\S\ref{agnsec}).  We then review
galaxies for which detections of both AGN or MBHs and NCs exist
in the literature (\S\ref{litsec}). We discuss the demographics of
galaxies with AGN and NCs, the relative masses of these CMOs in
galaxies where they co-exist, and the implications of this study for
CMO formation, in \S\ref{dissec}.  We conclude and discuss future
work in \S\ref{conclusions}.

\section{Sample Selection \& Properties} \label{samplesec}

We have used catalogs from multiple studies to create a sample of
nearby galaxies with known NCs.  This sample, which contains galaxies
of all Hubble types, is the starting point for finding galaxies that
contain both NCs and MBHs.

The Hubble Space Telescope (HST) has enabled NCs to be identified in a
large numbers of nearby galaxies with distances $\lesssim 30$~Mpc.
The use of HST is important both for distinguishing the NCs amidst the
crowded inner regions of galaxies and for resolving the clusters.  In
the studies used in our sample, the NCs have been selected from
broadband optical or NIR images, and are seen as compact sources
distinct from the underlying galaxy profile.  In the vast majority of
the selected galaxies, the NCs have been resolved, suggesting that
they are stellar sources and not AGN emission \citep[see discussion
in][]{rossa06}.  Spectral studies of a number of the NCs in our sample
confirm that their optical spectrum is dominated by starlight
\citep{walcher06,rossa06}.

We use the following catalogs of galaxies with nuclear star clusters:
\begin{enumerate}
\item Elliptical and lenticular galaxies from the ACS Virgo Cluster
Survey of \citet{cote06}.  This survey includes both giant and dwarf
elliptical galaxies in Virgo.  Of the 100 elliptical galaxies in the
survey, 51 contain NCs with measured properties \citep[type ``Ia''
nuclei in Table~1 of][]{cote06}.  These galaxies are all fainter than
$M_B \gtrsim -19$; some brighter galaxies have apparent NCs but
\citet{cote06} are unable to derive their properties.  The brightest
galaxies in their sample ($M_B < -20.5$) do not have any NCs.  Of the
51 NCs, five are unresolved (see Table~\ref{sampletab}).
\item Early-type spiral galaxies from
\citet{carollo97,carollo98,carollo02}.  Targets for their WFPC2 and
NICMOS snapshot programs are Sa-Sbc galaxies with $v_{hel} < 2500$
km/sec, angular diameter $>1$', and an inclination $<75^\circ$.
Combining their WFPC2 and NICMOS studies, there are a total of 58 out
of 94 galaxies with NCs; these are resolved in all but two cases.
\item Late-type spiral galaxies from \citet{boker02}.  This sample
includes bulgeless spiral galaxies of type Scd-Sm, with $v_{hel} <
2000$ km/sec and inclinations of $\lesssim30^\circ$.  Of 73
galaxies, 59 were found to have NCs.
\item Edge-on late-type (Sbc-Scd) galaxies from \citet{seth06}.  Of 
the 14 galaxies in this sample, 9 have NCs, 6 of which
are well resolved.
\end{enumerate}
 
Our sample includes 176 galaxies with nuclear star clusters spanning
all galaxy types.  Although the galaxy samples used are not complete,
none of the selection criteria for these samples depends on the
nuclear properties of the galaxies.

\begin{figure}
\plotone{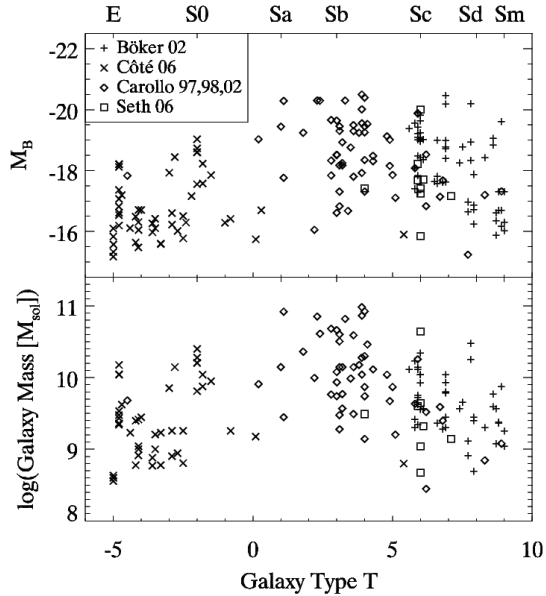}
\caption{{\it Top --} The galaxy type and absolute magnitude of all
176 of our sample galaxies with known nuclear star clusters.  The x-axis
gives the numerical galaxy type ``T'' and corresponding Hubble type
from Hyperleda \citep{paturel03}.  Symbols indicate the source catalog 
for each galaxy as discussed in the text.  {\it Bottom --} The galaxy mass 
for all 147 galaxies for which colors were available from Hyperleda.}
\label{galpropfig}
\end{figure}

Galaxy distances were determined from a variety of sources, including
using Virgo-infall corrected velocities from Hyperleda
\citep{paturel03} and
NED1D\footnote{http://nedwww.ipac.caltech.edu/level5/NED1D/} for other
distance indicators.  For the Virgo cluster galaxies, a distance of
16.5 Mpc was assumed \citep{tonry01,mei07}.  The galaxies range in
distance between 2 and 40 Mpc, with most having distances between 10
and 30~Mpc.  Galaxy magnitudes and Hubble types (including numerical
types, ``T'') were also determined from Hyperleda.
Figure~\ref{galpropfig} shows the type and absolute magnitude of all
176 galaxies in our sample.  The B-band absolute magnitudes of these
galaxies range between $-15$ and $-21$.  Galaxy masses were obtained
using galaxy colors to estimate the M/L ratios from \citet{bell03}.
We obtained optical color estimates for 147 of the 176 galaxies from
Hyperleda \citep{paturel03}, including their B-V colors, as well as
colors from the Sloan Digital Sky Survey \citep[SDSS;][]{york00} and
6dF survey \citep{jones04}.  To obtain masses we used both the
Hyperleda total B magnitudes corrected for internal and foreground
extinction (also obtained from Hyperleda), and where available, K band
magnitudes from 2MASS and DENIS.  In cases where multiple sets of
photometric data were available, the median value for the galaxy mass
was used.
The masses determined for a single galaxy using different methods
typically differ by $\sim$35\% (0.15 dex).  The sample galaxy
properties and distances are given in Table~\ref{sampletab}.

\subsection{Nuclear star cluster masses}

The NCs in our sample have magnitudes measured in many different
bands.  To compare the NC properties for the entire sample we
therefore estimated the mass for each NC.  There is
strong evidence that NCs, at least in spiral galaxies, have
complicated star-formation histories \citep{walcher06,rossa06,seth06}.
Therefore, derivation of NC masses from integrated
magnitudes is not straightforward.  We have estimated the masses for
the NCs in our sample using a variety of methods, as detailed below.  The
derived masses are shown in Table~\ref{sampletab} and are used in
determining the relative masses of NCs and MBHs and for examining the
demographics of galaxies hosting both types of objects (see
\S\ref{dissec}).

\begin{enumerate}
\item The best available mass estimates are dynamical
measurements of 9 NCs in the B\"oker sample by
\citet{walcher05} and masses from population synthesis fits for an
additional 15 spiral galaxies in \citet{rossa06}.  

\item For the \citet{cote06} clusters, we followed the prescription of
\citet{ferrarese06}, who estimate the NC mass by assuming an age of
5~Gyr and use the cluster's published $g-z$ color to determine the
metallicity and thus the appropriate mass-to-light (M/L) ratio from
\citet{bruzual03}.

\item For the early type spiral galaxies in the Carollo sample, we use
the mean B-band M/L ratio of $3.64\pm1.03$ derived for early type
spiral NCs by \citet{rossa06}.  For most of the galaxies, both V
(WFPC2-F606W) and H (NICMOS-F160) magnitudes are available for each
cluster, while a minority of galaxies have just one or the other
magnitude available.  We derive B-band magnitudes by assuming the
colors of an SSP with Z $=0.030$ \citep[matching the derived mean
metallicity in][]{rossa06} and age of 5.9 Gyr to match the B-band M/L
ratio \citep{bruzual03}.  These magnitudes were corrected for the
foreground and estimated internal extinction as determined from
Hyperleda \citep{paturel03}, the mean extinction correction was $A_{V}
= 0.27$.

\item For the late-type spiral galaxies, \citet{walcher05} has shown
that the typical I-band M/L ratio is $0.50\pm0.37$.  We used this to
derive masses from the I-band magnitudes in \citet{boker02} and
\citet{seth06}, after correcting for foreground and internal
extinction.  We note the possibility that because of the selection of
objects with bright apparent magnitudes, the \citet{walcher05}
spectroscopic sample may not be representative of the \citet{boker02}
sample as a whole.  This may result in an underestimate of the M/L
ratio, as the study would favor younger, brighter NCs.

\end{enumerate}

For the NCs with available dynamical masses \citep{walcher05} or
stellar population model masses \citep{rossa06}, the
agreement between these masses and those obtained using the methods outlined above is good,
with a mean difference of $-0.02\pm0.34$ dex.  This standard deviation
of 0.34 dex (factor of $\sim2$) gives some indication of the error in
our NC mass determinations.

\begin{figure}
\plotone{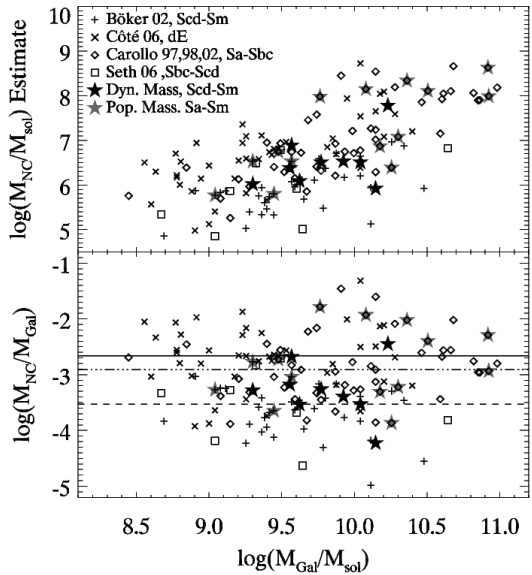}
\caption{{\it Top --} Nuclear star clusters mass vs.~galaxy mass for
all the clusters with derived galaxy masses in our sample.  {\it
Bottom --} ratio of nuclear star cluster mass to galaxy mass.
Overplotted are the median ratios for the elliptical galaxies (solid
line), the early type spirals (dot-dashed line) and the late-type
spirals (dashed-line). In both panels the stars indicate dynamical
masses measured for late-type spirals by \citet{walcher05} (black
stars), and the spectral synthesis masses from \citet{rossa06} (gray
stars).}
\label{nucmassfig}
\end{figure}

Figure~\ref{nucmassfig} shows the derived NC masses as a function of
galaxy mass.  Although correlations between NC and galaxy luminosities or
masses have been shown for samples of galaxies of a single type
\citep{carollo98,boker04a,cote06,rossa06,ferrarese06,wehner06}, this is
the first time they have been compared across all Hubble types.
Figure~\ref{nucmassfig} shows the expected correlation between NC mass
and galaxy mass.  However, we find an offset between earlier- and
later-type galaxies, with the later-type galaxies having less massive
NCs at a given galaxy mass.  This can be seen in the bottom panel of
Figure~\ref{nucmassfig}, where we plot the ratio of NC to galaxy mass.
Lines indicating the median NC to galaxy mass ratio for elliptical,
early type spiral, and late-type spiral galaxies show that late-type
spirals have NC masses about an order of magnitude below
elliptical galaxies of the same mass.  We will discuss these results
in greater detail in a future paper.

\section{Presence of AGN} \label{agnsec}

In this section we analyze the evidence for active galactic nuclei
(AGN) in our sample galaxies using optical spectroscopy
(\S\ref{agnsec}.1) and radio and X-ray data (\S\ref{agnsec}.2). We then
examine the strength of the evidence for these detections being
massive black holes in \S\ref{agnsec}.3.  

\subsection{Emission-line spectroscopy}

\begin{figure*}
\plotone{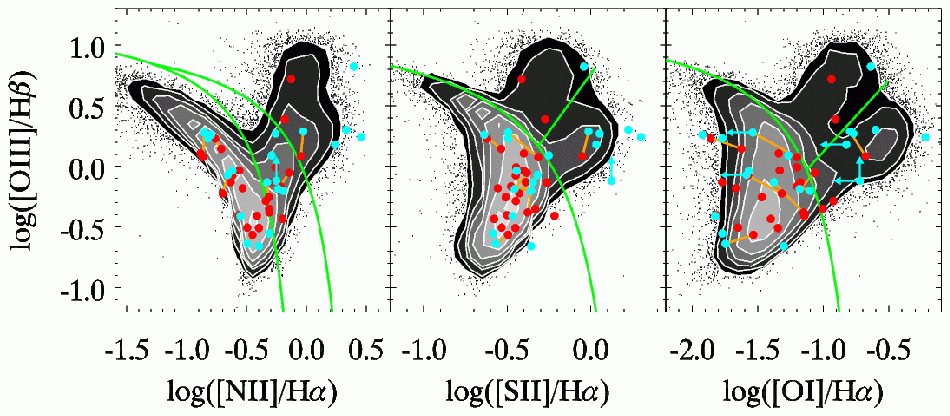}
\caption{Emission line diagrams for galaxies in our sample.  Red
points show line ratios derived from SDSS DR6 data, blue are from
\citet{ho97a}; orange lines connect galaxies with both types of
spectra. Grayscale and small black points indicate the Garching
reduction of the full SDSS DR4 sample. Green lines indicate the
classification system adopted from \citet{kewley06}.}
\label{bptfig}
\end{figure*}

Of the 176 galaxies in our sample, 70 had possible nuclear spectra
available in the SDSS Data Release 6 (DR6) \citep{sloandr6}.  After
visually inspecting the location of each spectrum, we found that 62
spectra were coincident with the galaxy nuclei.  Emission line fluxes
determined after modelling of the underlying stellar populations were
kindly provided by C. Tremonti ({\it private communication}) using the
method described in \citet{tremonti04}, and available for SDSS Data
Release 4 (DR4) data at {\tt
http://www.mpa-garching.mpg.de/SDSS/DR4/}.  The DR6 spectra used
differ somewhat from earlier spectra.  Most notably, the
spectrophotometric zeropoint has changed,
increasing fluxes by $\sim35\%$ but leaving line ratios unchanged.  In
general, the line ratios from the DR6 spectra agree very well with
those in DR4.  The formal errors on the line fluxes were scaled by
factors of $1.4-2.5$ (depending on the line) to include errors in the
continuum subtraction and flux calibration derived from sources with
multiple spectra as described on the Garching DR4 
website\footnote{http://www.mpa-garching.mpg.de/SDSS/DR4/raw\_data.html}.
These errors were then propagated to the line ratios.  We select
emission line galaxies in the SDSS spectrum by requiring that three
of four strong lines (H$\beta$, [OIII] $\lambda$5007, H$\alpha$, and
[NII] $\lambda$6584) have detections above $3\sigma$.  Of the 62
spectra in the sample, 25 meet this criterion.

We also find 23 galaxies that were observed as part of the Palomar
Survey \citep{filippenko85}, 20 of which have detected emission lines
\citep[][hereafter HFS97\footnote{We use the emission line
measurements from HFS97, but not their spectral
classifications.}]{ho97a}.  As with the SDSS data used above, the
emission line measurements are made after subtraction of the
underlying stellar population based on spectral modelling.  Errors in
the line ratios were estimated using the data quality flags and
assuming a conservative baseline uncertainty of 30\%, with 50\% and
100\% uncertainties for sources with uncertainty flags of `b' and `c'
respectively.  Ten of these galaxies overlap with the SDSS spectra,
giving a total of 75 galaxies for which we have nuclear spectra.

Of the 75 galaxies for which we have nuclear spectra, 39 have weak or
undetected emission lines; we classify these galaxies as having
absorption-dominated spectra.  For the remaining 36 emission-line
spectra, we followed the classification scheme of \citet{kewley06} to
separate the sample into star-forming galaxies, composite objects, or
Seyfert and LINER AGN.  This classification scheme relies on four line
ratios, [OIII]/H$\beta$, [OI]/H$\alpha$, [NII]/H$\alpha$, and
[SII]/H$\alpha$.  

Three emission line ratio diagrams \citep[BPT diagrams;][]{baldwin81}
are shown in Figure~\ref{bptfig}.  This figure shows line ratios from
the full DR4 sample of galaxies in grayscale contours and small black
points along with our sample galaxies as red (SDSS) and blue (HFS97) dots.
The green lines indicate the demarcation lines used in the
classification scheme \citep{kewley06}. The primary classification of
galaxies into star-forming, composite objects, or AGN is done using
the [NII]/H$\alpha$ vs.~[OIII]/H$\beta$ diagram (left panel of
Figure~\ref{bptfig})\footnote{Due to the proximity of the emission
lines used in these ratios, the [NII]/H$\alpha$ vs.~[OIII]/H$\beta$
diagram is remarkably insensitive to reddening: 5 magnitudes of
reddening creates a change of 0.006 and 0.08 dex in the
[NII]/H$\alpha$ and [OIII]/H$\beta$ ratios respectively.}. Spectra are
classified as being AGN if they fall in the upper right part of the
diagram above the theoretical maximum starburst line of
\citet{kewley01} in the [NII]/H$\alpha$ vs.~[OIII]/H$\beta$ diagram.
Composite objects have emission lines thought to be caused by a mix of
AGN and star-forming lines, and have line ratios falling below the
\citet{kewley01} line and above the empirical \citet{kauffmann03}
line.  Galaxies below the \citet{kauffmann03} line have line ratios
dominated by star formation.  Further separation of AGN into LINER and
Seyfert galaxies is done using the [SII]/H$\alpha$ vs.~[OIII]/H$\beta$
and [OI]/H$\alpha$ vs.~[OIII]/H$\beta$ diagrams (right two panels of
Figure~\ref{bptfig}), with Seyferts lying at higher values of
[OIII]/H$\beta$.

Using the [NII]/H$\alpha$ vs.~[OIII]/H$\beta$ diagram, we classify 18
galaxies as having star-forming (``{\rm HII}'') spectra, 11 as having
composite spectra (``{\rm C}''), and 7 as having AGN-like spectra
(``{\rm AGN}'').
The other line ratio diagrams suggest that 3 of the AGN are Seyferts
(``S2'') and 4 are LINERS (``L2''), while an additional 3 composite
galaxies are also found to have LINER-like line ratios.  For the ten
overlapping galaxies (connected by orange lines in Figure~\ref{bptfig}),
the classifications between the HFS97 and SDSS data agree in all
cases, except for NGC~5879. This galaxy is classified as an AGN from both
spectra, but is found to be a LINER only from the SDSS spectrum.

Our classifications also agree well with the somewhat different
classification system used by HFS97 for the Palomar
galaxies\footnote{The HFS97 classification system gives a primary role
to the [OI]/H$\alpha$ line ratio, which is quite weak in many of our
spectra and significantly more susceptible to reddening than the
[NII]/H$\alpha$ ratio.}.  
Eleven of the eighteen sources we classified as composite or
AGN are in the HFS97 sample.  Of these, 8 are classified as transition
or AGN by HFS97.  Specifically, the five galaxies that we classify as
LINER and Seyfert AGN match their classifications exactly, while for
the six galaxies we classified as composite, half of them were
classified as star-forming, with the other half being LINERS or
transition objects in their classification scheme.  Both the composite
class from \citet{kewley06} and the transition class from HFS97 are
thought to result from a mix of star-forming and AGN spectra, thus a
correspondence between these classes is expected.  In the HFS97
classification the transition objects are objects with a mix of LINER
and star-forming spectra, while in our classification they can be a
mix of LINER or Seyfert and star-forming spectra.

One of the galaxies in the HFS97 sample, NGC~4750, was found to have
broad H$\alpha$ emission, and is thus classified by them as a L1.9.
We have adopted this classification.  
Also, \citet{shields08} has
recently found broad [NII] (but not H$\alpha$) emission lines in the
composite galaxy NGC~1042.  Visual inspection of the model-subtracted
SDSS spectra suggest that none has any obvious broad emission lines.

The line ratios used in our classification are given in
Table~\ref{hiitab} for the star-forming galaxies and in
Table~\ref{agntab} for the composite and AGN galaxies (C/AGN).  For
the star-forming nuclei, Table~\ref{hiitab} also gives the
star-formation rates based on H$\alpha$ luminosities derived from the
relation given in \citet{kennicutt98a}. Table~\ref{agntab} gives the
[OIII] luminosities for the C/AGN nuclei which are used in
\S\ref{dissec}.2.  These line luminosities have been corrected for
reddening assuming H$\alpha$/H$\beta$=2.85 for the star-forming
galaxies and H$\alpha$/H$\beta$=3.1 for the C/AGN galaxies
\citep{osterbrock89,kewley06}.

To get a sense of the uncertainty in our classification, we repeated
the classification after both adding and subtracting the 1$\sigma$
errors to the [NII]/H$\alpha$ and [OIII]/H$\beta$ ratios.  This
changes the classification for a handful of galaxies: the number of
composite galaxies varies between 8 and 12, while the number of AGN
varies between 6 and 8.  Seven galaxies have error bars crossing the
HII/C boundary: C-classified NGC~428, 3423, 4206, 4517, and 4625 and
HII-classified NGC~2964 and VCC~1250.  Two galaxies have error bars
crossing the AGN/C boundary: C-classified galaxy VCC~1619 and
AGN-classified NGC~4411B.

Although the galaxies in our sample are quite nearby, the physical
resolution of the spectra (3'' fibers for SDSS, 2'' slit for HFS97) is
still significantly larger than the typical cluster sizes.  For a
galaxy at the median distance of our sample (16.5~Mpc), the 
corresponding spatial
resolution is $\sim200$~pc.  This could lead to detection of
star-formation not coincident with the nucleus \citep{shields07}, and
might be expected to dilute weak AGN emission, therefore causing
genuine AGN to be classified as composite objects.  

In summary, from the 75 galaxies with available optical spectra, we
find 18 that have composite or AGN spectra.  We discuss the fraction
of these galaxies that have MBHs in \S\ref{agnsec}.3.  We now discuss
the radio and X-ray properties of our sample galaxies. 

\subsection{Radio and X-ray correlations}
In order to explore the multiwavelength properties of the galaxies in
our sample, we matched the NC positions to a number of radio and X-ray
catalogs. While the cataloged observations at these wavelengths are
unable to resolve structures on the sub-arcsecond scale of the NCs,
detections in the radio and/or the X-ray regimes can be used to place
limits on the emission from putative AGN independently of the
information derived from optical spectra.

\subsubsection{Radio data}

Galaxies that are radio sources are thought to host either active star
formation or AGN.  
Radio observation of AGN show they have compact nuclear radio sources
with high brightness temperatures that cannot be reproduced by
starbursts \citep[e.g.,][]{terashima03}.  Such radio sources are detected
in about a third of galaxies classified as composite or AGN objects
in the Palomar survey \citep{nagar05}.  Sources with $L_{1.4\ GHz} \ge
10^{23}$~W~Hz$^{-1}$ are generally called radio-loud AGN, and are
unambiguous evidence of an AGN \citep{best04}.  However, these sources
are typically associated with very massive galaxies, and thus it is
unlikely we will find radio-loud AGN in our sample \citep{croft07}.

We used the Very Large Array (VLA) Faint Images of the Radio Sky at
Twenty-Centimeters \citep[FIRST;][]{first} to search for radio
counterparts to the galaxies in our sample. FIRST is the deepest
large-scale radio survey currently available; the limiting flux
density is about $1.0$ mJy, the survey resolution is $\sim5\asec$, and
the footprint is roughly the same as that of SDSS.
We queried FIRST for radio sources within $30''$ of our NCs and found
$13$ matches\footnote{Matching to the National Radio Astronomy
Observatory (NRAO) VLA Sky Survey \citep[NVSS;][]{nvss}, which covers
the sky north of $\delta = -40^\circ$ and includes sources stronger
than $\sim2.5$ mJy, returned a larger number of matches, $52$,
including all of the FIRST-detected objects. However, the resolution
for NVSS is roughly $45\asec$ FWHM and the survey data are therefore
far less useful for our purposes.}. The median separation between the
radio source and the cluster position is $3\pm9\asec$ (see
Table~\ref{radio_data}). Of the $13$ matched galaxies, we have
optical-based classifications for nine
(see Table~\ref{radio_data}). One galaxy, NGC~5377, is classified as
an AGN/L2, while three others are classified as composite objects in
our analysis. The remaining five galaxies have star-forming optical
spectra.  Unsurprisingly, the $L_{1.4 GHz}$ for all these objects is
well below $10^{23}$ W Hz$^{-1}$.

\begin{figure}
\includegraphics[width=.99\columnwidth]{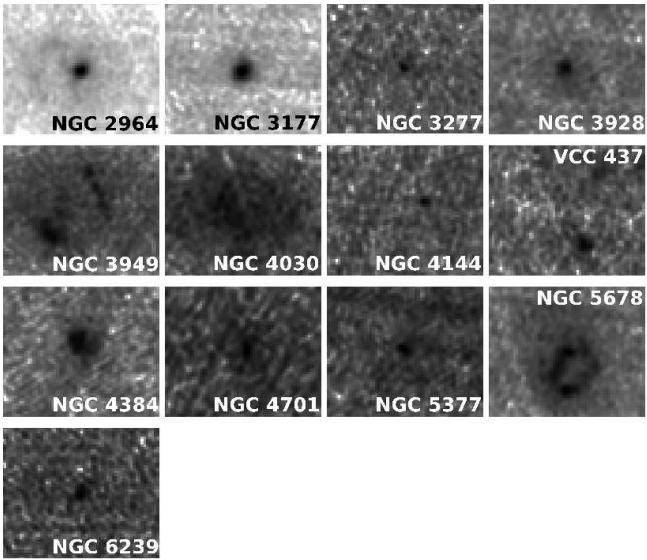}
\caption{FIRST images of the $13$ galaxies for which we find a catalog
match within $30''$. Each image is roughly $1.6'$ in height and $2'$
in width and centered on the optical position of the nuclear star cluster
in the galaxy.}
\label{FIRST_cutouts} 
\end{figure}

Visual inspection of the corresponding radio images did not find
morphological evidence for the presence of an AGN (e.g., jets) in any
of these (see Figure~\ref{FIRST_cutouts}). In most cases, the radio
emission appears diffuse, as expected if it is due to star
formation. However, NGC~5377 and two of the three composite objects
(NGC~3177 and NGC~3928) with radio detections appear as point sources in
the FIRST images, consistent with the presence of a possible nuclear
AGN source.  For NGC~5377, higher resolution observations do confirm
this source as an AGN \citep{nagar05}; similar radio
observations are needed to determine the nature of the other sources.  

An additional $94$ galaxies from our sample fall within the FIRST
footprint but do not have detections. \citet{white07} have developed a
method for obtaining the FIRST radio flux density for a group of
sources when individual group members are undetected in the
survey. Accordingly, we calculated a stacked image for our undetected
sources. We applied the correction, prescribed in \citet{white07}, to
account for the snapshot bias and derived our errors from the
bootstrapping method \citep{dasboot}, which tests how individual
entries affect the stacked average.
We found that the average radio flux density for the undetected
galaxies in the FIRST footprint is $\sim 90 \pm 20\ \mu$Jy (whereas
the detected galaxies have flux densities of 1.6 to 68 mJy). 
This stacked ``detection'' provides an average measurement of the
radio intensity for the typical nuclear region in these galaxies. At
best, this suggests that any putative radio AGN in these galaxies is
extremely weak.  
We note that among these non-detections are four galaxies we have
classified as C/AGN objects using optical spectroscopy: NGC~4411B,
VCC~1619, NGC~5806, and NGC~5879.

\subsubsection{X-ray data}
X-rays are one of the most direct evidences of nuclear activity and
X-ray observations are therefore essential in revealing the accretion
processes taking place near the central black holes in AGN. With the
spatial resolution of the new generation of X-ray telescopes, it is
becoming possible to isolate low-luminosity AGN from other X-ray
sources in a galaxy \citep[e.g.,][]{ho01}. In particular, the ability
to detect photons with energies of several keV allows for the
uncovering of AGN hidden at other wavelengths by column densities as
high as $N_H \sim 10^{24}$ cm$^{-2}$ \citep{ho01}. 

We matched our sample to the combined {\it ROSAT} All-Sky Survey
(RASS) Bright and Faint Source Catalogs \citep[][]{voges99, fsc}. We
also queried the High Energy Astrophysics Science Archive Research
Center website for matches from the various catalogs of pointed {\it
ROSAT} observations, {\it Chandra} sources, and {\it XMM-Newton}
detections\footnote{This research has made use of data obtained from
the High Energy Astrophysics Science Archive Research Center
(HEASARC), provided by NASA's Goddard Space Flight Center.}.

The X-ray data obtained in this fashion is rather heterogeneous, since
the queried catalogs and instruments are very different in nature. For
example, while none of the {\it Chandra} or {\it XMM} surveys covers
as much sky as the RASS, the latter's positional accuracy is generally
relatively poor (typically at least $15''$). In a similar vein, {\it
Chandra} and {\it XMM} both have greater sensitivity than {\it ROSAT},
but the telescopes are not designed to detect sources in exactly the
same energy ranges.

In order to produce the data for the matches listed in
Table~\ref{xray_data}, we proceeded as follows:

\begin{enumerate} 
\item Since {\it Chandra} typically has better positional accuracy
than {\it XMM}, and both typically have better accuracy than {\it
ROSAT}, we used {\it Chandra} data preferentially, then {\it XMM},
then {\it ROSAT}, in our analysis.

\item For galaxies with {\it Chandra} and/or {\it XMM} detections,
there were frequently multiple X-ray sources. In those cases we chose
the source closest to the optical position of the NC\footnote{The positional accuracy of our NC
positions is dominated by the uncertainty in the HST astrometry, which is typically accurate to within $1-2\asec$.}. The
median offset for the $13$ {\it Chandra}/{\it XMM} sources is
$1.3\asec\pm1.4\asec$, with the largest offset being
$4.9\asec$. All the {\it Chandra} data
come from XAssist \citep[an automated extraction pipeline for X-ray
data;][]{ptak03}; these are the sources with an ``X'' prefix in
Table~\ref{xray_data}. The {\it XMM} data are from the second catalog
of the {\it XMM} serendipitous survey (2XMM sources; Watson et al., in
prep.).

\item Of the $8$ galaxies with RASS matches, half had detections in
pointed catalogs, and we therefore used those data to characterize the
X-ray sources. Three of the other galaxies, NGC~3259, NGC~4030, and
NGC~4540, are published X-ray sources \citep[identified as such
by][]{veron04,moran96, mick06}. The positional offsets between the
RASS source and the NC are small ($\leq10\asec$) for the two first
galaxies, and the associations with the RASS sources seem secure.
However, NGC~4540 and NGC~2566 both have large positional
uncertainties and offsets from the NCs.  Pointed observations are
clearly required to confirm that these two galaxies are X-ray sources,
and we consider these associations to be tentative.

\item Four more galaxies are included in various {\it ROSAT} pointed
surveys and/or identified as X-ray sources in the literature. Two of
these are from catalogs of High Resolution Imager (HRI) sources
\citep{panzera03,rosat00}\footnote{The \citet{panzera03} catalog is a
re-analysis of the HRI observations with exposures longer than $100$
s; the other catalog includes all HRI pointed observations.}. In both
these galaxies, NGC~1385 and NGC~6000, the offset with the NC is less
than the $5\asec$ nominal positional accuracy for HRI observations
\citep{flesch04}. The two other sources were detected with the other
X-ray instrument aboard {\it ROSAT}, the Position Sensitive
Proportional Counter (PSPC), whose positional accuracy is closer to
$30\asec$. \citet{flesch04} identify NGC~600 as a
PSPC source with $61\%$ confidence\footnote{\citet{flesch04} use the
\citet[][]{rosat00} catalog of PSPC pointed observations as the basis
for this match.}; the offset between the X-ray source (positional
uncertainty $13\asec$) and the NC is $10\asec$. \citet{white96}
include NGC~3445 in their WGACAT\footnote{\citet[][]{white96}
generated their own point source catalog from all publicly available
{\it ROSAT} PSPC observations.}.

\item We used WebPIMMS \citep{mukai93} to calculate $2 - 10$ keV
fluxes for all of these sources, assuming a canonical intrinsic
power-law spectrum with photon index $\Gamma = 1.8$ \citep[for
low-luminosity AGN $\Gamma$ ranges between $1.6$ and
$2.0$;][]{terashima03} absorbed to the Galactic value. We then
calculate the X-ray luminosities given in column 7 of
Table~\ref{xray_data}. These values are consistent with those in the
literature for the known X-ray-emitting galaxies, once differences in
the adopted distances and energy bands are taken into account.
\end{enumerate}

A complicating factor in interpreting the X-ray luminosities we obtain
at the end of this process is the differing point spread function
(PSF) for each of these telescopes. While, broadly speaking, {\it
Chandra}, {\it XMM}, and the HRI on {\it ROSAT} have similar PSFs
(from $\sim1$ to $6\asec$), the {\it ROSAT} PSPC has a PSF with a
$\sim30\asec$ FWHM \citep{panzera03}, and the PSFs all vary with
position on the detector and photon energy. Determining which sources
are extended and which are truly point-like is therefore difficult,
with the exception of the {\it Chandra} sources, for which XAssist
provides sub-arcsecond measurements of source extent (listed in column
6 of Table~\ref{xray_data}). The situation with the 2XMM sources is
relatively straightforward: any source that is smaller than the
$6\asec$ PSF and hence unresolved is set in the catalog to have a
$0\asec$ extent.

The RASS sources (PSPC detections) all have associated cataloged
extent measurements and extent likelihoods \citep[see][]{voges99};
however, these values should be treated with caution (e.g., for bright
sources the deviations of the PSF from a Gaussian lead to incorrect
extent measurements), and are considered reliable only if the extent
measurement is $>10\asec$ and the likelihood is $>10$ \citep[F.
Haberl, {\it private communication}; see also][]{haberl00}. For three
of the RASS sources the measured extent and extent likelihood are both
$0\asec$; for the fourth, RX J120023$-$01055 (NGC~4030), the cataloged
extent is $9\asec$, but the extent likelihood is $1$, implying the
source is actually unresolved.

For the sources with pointed {\it ROSAT} observational data, the
definition of source extension varies from catalog to catalog. For the
two sources detected by the HRI, 1BMW 033728.0$-$243003 and 1RXH
J154949.7$-$292310 (counterparts to NGC~1385 and NGC~6000), the
extents are given as $0\asec$. However, \citet{panzera03} give the
actual source extent for NGC~1385 as $14\asec$ (with extent likelihood
$0$). The catalog entry for NGC~6000 indicates that it is not
extended, but does not give an associated measurement. As for the PSPC
sources, 2RXP J013305.7$-$071835 (NGC~600) is fit by a Gaussian with
$\sigma = 12\asec$, and is given an extent likelihood of $0$ in the
PSPC catalog \citep{rosat00}. Finally, the WGACAT is defined as a
point-source catalog, and no measurements are included in the
available data.

In summary, we find a total of 22 X-ray sources associated with NCs in
our sample.  We evaluate how many of these may possess black holes in
the following section.

\subsection{Do these sources harbor black holes?}

We now consider how reliably we can infer the presence of MBHs in the
galaxies for which our multiwavelength data indicate the presence of
AGN.

For the optical data, considerable work has been done to determine
whether low-luminosity Seyferts, LINERS, and transition/composite
objects do in fact represent MBH accretion.  While Seyfert line ratios
are a strong indicator of MBH accretion, it is not clear whether the
same is true for all LINERs and transition/composite objects
\citep[see review by][]{ho04}.  Several recent studies of LINER
galaxies suggest that a majority of them do in fact indicate the
presence of an MBH based on high resolution radio observations
\citep{nagar05}, X-ray emission \citep{dudik05,gonzalezmartin06}, and
UV variability \citep{maoz05}.  There are cases where LINER line
ratios do seem to be produced by star formation. For instance, in the
LINER galaxy M61 (NGC~4303), the recently formed stars provide more
than enough ionizing flux to explain the observed H$\alpha$ emission
\citep{colina02}, suggesting that LINERS may sometimes just be caused
by bursts of star formation.  However, we note that in the context of
our sample, a majority of the LINERS are found in early type spiral
galaxies, whose NCs are known to only rarely host significant young
stellar populations \citep{rossa06}.  The nature of composite or
transition objects, proposed to be a mix of AGN emission and star
formation, is even more difficult to determine \citep{ho04}.  From a
radio survey of the HFS97 catalog, \citet{nagar05} finds 16\% of
transition objects have compact radio detections, compared to
$\sim$45\% for the Seyferts and LINERs.  
The observation of broad-emission lines in NGC~1042 (which we classify
as a composite object) by \citet{shields08} does provide strong
evidence for a MBH in that galaxy.  In summary, the presence
of an AGN implies the presence of MBHs reliably in Seyferts and a
majority of LINERS, but less reliably for sources with composite
spectra.

For the radio data, the presence of AGN can only be safely inferred in
cases where the emission is above a certain threshold. Below that
threshold \citep[typically $\sim$10$^{23}$~W~Hz$^{-1}$;][]{best04},
either star-formation or low-level AGN can be responsible for the
radio emission. The $13$ galaxies with FIRST detections in our sample
all have luminosities $<10^{22}$~W~Hz$^{-1}$, which is unsurprising
given the low masses of galaxies in our sample and the observed
correlation between galaxy mass and radio-loudness \citep{croft07}. We
therefore cannot use $L_R$ alone to identify radio AGN in our sample.
While morphological information could distinguish between
star-formation and AGN emission in galaxies with lower radio
luminosities, the resolution of FIRST data is insufficient to place
strong constraints on the nature of the sources in these galaxies. The
radio data are therefore useful only in setting upper limits on any
(low-level) emission from MBHs in these galaxies, which require higher
resolution observations to be detected with confidence.

For the X-ray data, which provide insight into a larger fraction of
the AGN bolometric flux than the narrow optical emission lines, any
well localized source with $L_X\gtrsim2\times10^{39}$~ergs/sec very
likely originates from an MBH \citep[e.g.][]{vandermarel04}.  This
criterion is met by only four of our sources, NGC~4750, NGC~6000,
NGC~6951, and NGC~7418 (Table~\ref{xray_data}), two of which (NGC~4750
and NGC~6951) have available optical spectra and are classified as
AGN.  The other sources are either luminous but poorly resolved {\it
ROSAT}/PSPC detections, or lower luminosity ($\sim10^{38}$~ergs/sec)
{\it Chandra} or {\it XMM} detections.  For the {\it ROSAT}/PSPC
detections, the resolution is quite poor (FWHM $\sim$30''), and thus
covers a significant portion of the galaxy ($\sim$3~kpc).  The X-ray
fluxes are within the range expected for normal galaxies
\citep{shapley01}, and thus could originate either from collections of
other X-ray sources or accretion onto an MBH (note we include the
extended {\it ROSAT}/HRI source in NGC~1385 in this category as well).
For the higher resolution {\it Chandra} and {\it XMM} observations,
which are localized to be coincident with the NC (see
Table~\ref{xray_data}), the sources with low X-ray luminosities
overlap with those expected for X-ray binaries \citep[][]{fabbiano06}.
For instance, luminous X-ray sources in the NCs of M33 and NGC~2403
are both thought to be X-ray binaries \citep{dubus04,yukita07}.  Also,
low-mass X-ray binaries (LMXBs) are commonly found in massive
clusters; in a survey of globular clusters in Virgo, $\sim$25$\%$ of
the brightest globular clusters were found to have LMXBs
\citep{sivakoff07}.  The {\it Chandra} and {\it XMM} detections with
$L_X \lesssim 10^{39}$~ergs/sec are therefore consistent with the
presence of X-ray binaries in these NCs.  However, many low-luminosity
AGN appear to have X-ray fluxes in this luminosity range as well
\citep[e.g.,][]{dudik05,panessa06}.  In conclusion, the presence of an
MBH powered source is strongly indicated for four X-ray sources, while
for the remainder of the sources the X-ray energy could be resulting
either from MBH accretion or from other X-ray sources.

Despite these uncertainties in identifying the presence of MBHs in
single galaxies in our sample, the data presented here
collectively provide good evidence that the overlap between NCs and MBHs is common.
In total, 10 galaxies in our sample show strong evidence of an MBH, with
$\sim$30 more having some indication of a possible MBH.  Furthermore,
it is likely that many MBHs in our sample galaxies do not have enough
activity to bring them to our attention; for example neither the
central black hole of the Milky Way \citep{narayan98,shields07} nor
that of M32 \citep{ho03a}, galaxies similar to those in our sample,
would likely be detected as AGN in our study.

\begin{figure*}
\plottwo{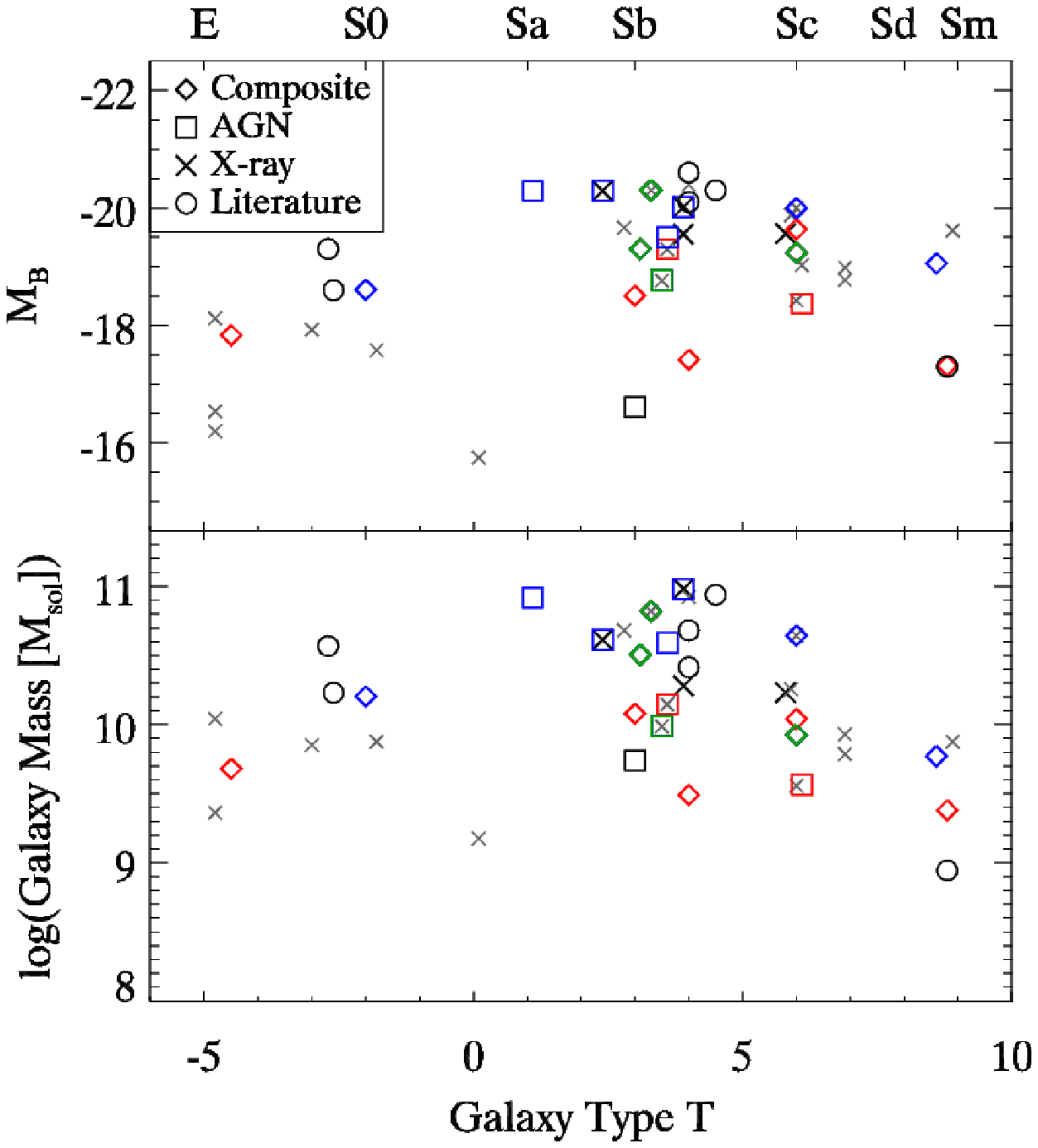}{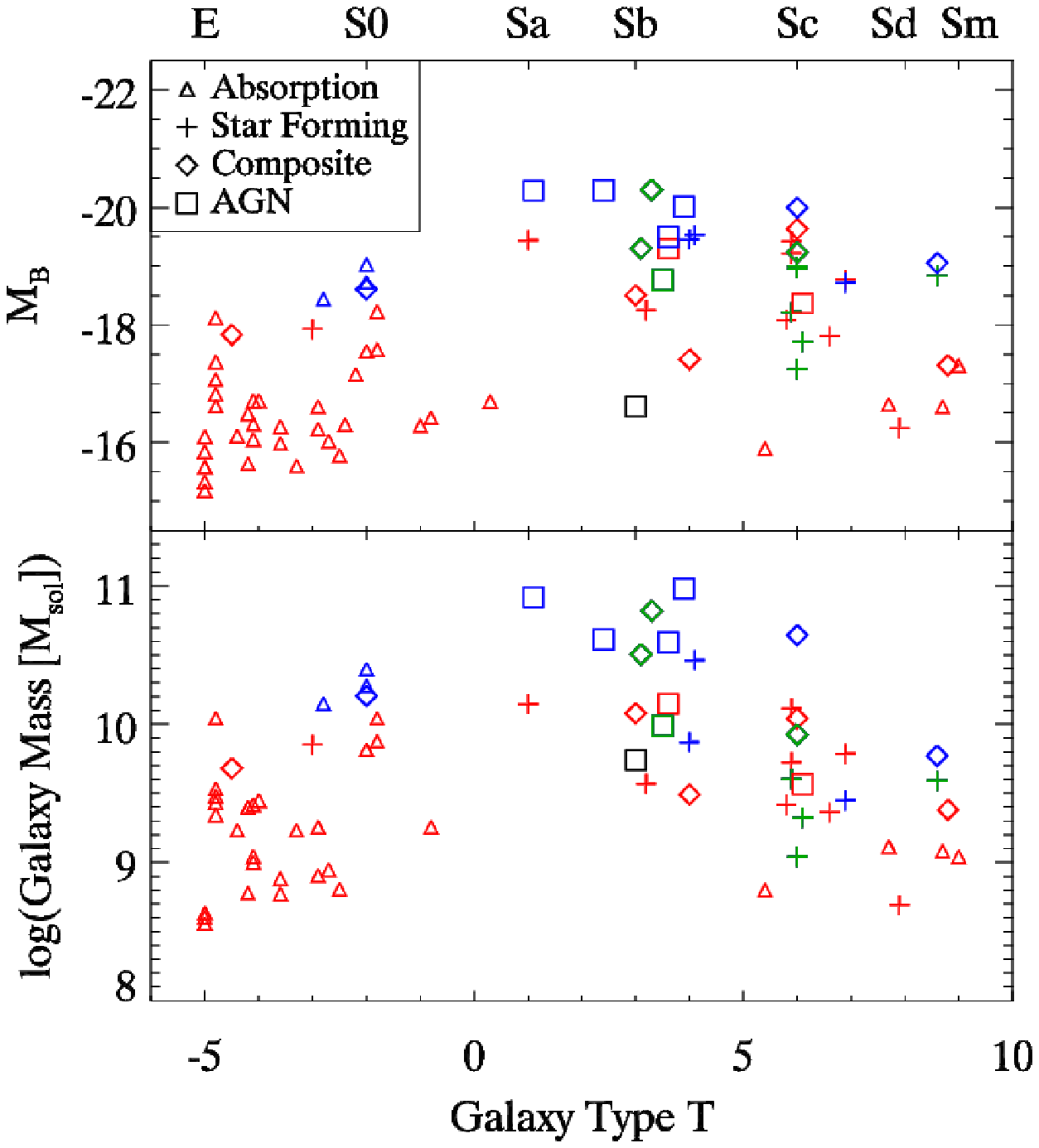}
\caption{{\it Left --} The galaxy properties of all galaxies with NCs
that also host candidate AGN.  The red points indicate spectra from
HFS97, blue points are SDSS spectra, and green points have both HFS97
and SDSS spectra.  The four X-ray sources that are very likely AGN
(see \S\ref{agnsec}.3) are shown with large black X's, the rest of the
detections have smaller gray X's.  The six sources from the literature
not contained within our sample are shown with circles
(\S\ref{litsec}).  The black square is ESO205-G7 \citep[see
\S\ref{litsec},][]{rossa06}.  {\it Right --} The galaxy type and
absolute magnitude of all galaxies with spectra, colored as in the
left panel.  Note that because mass estimates are not available for
all the galaxies (see \S\ref{samplesec}), some data points in the top
panel are missing in the bottom panel.}
\label{galagnfig}
\end{figure*}

\section{Previously known AGN/NC sources} \label{litsec}

In this section we present galaxies previously known to host both a NC
and an MBH or AGN.  These can be added to our sample to explore the types
of galaxies in which NCs and MBHs overlap.  Furthermore, four
galaxies presented below have measured black hole masses, enabling a
comparison of their NC and MBH mass (\S\ref{dissec}.2).

We find nine galaxies previously identified as having both NCs and an
AGN or MBH, three of which are contained in our sample.  These are all
galaxies in which the presence of both a NC and MBH/AGN is well
substantiated \citep[but see also][]{ghosh06,decarli07}.

The Milky Way hosts both a NC and an MBH.  The NC in the Milky Way was
first described by \citet{becklin68} using IR observations, and was
found to have a FWHM of $3-5\amin$ ($\sim$10~pc), consistent with NCs in
similar galaxies \citep{carollo02}.  More recent studies using both
star counts and kinematics show that the mass enclosed within the
central 10~pc is $\sim3\times10^7$~M$_\odot$
\citep{genzel96,schodel07}.  The Galactic MBH has a mass of
$3.7\times10^6$~M$_\odot$ \citep[][]{ghez05}.  

Another well known galaxy with a NC, AGN, and MBH is NGC~4395, an Sm
type galaxy at a distance of 4.3~Mpc \citep{thim04} with $M_B=-17.30$
\citep{paturel03}.  The NC in NGC~4395 \citep{matthews99} has an
$M_I=-11.3$, an effective radius of $0.19\asec$ (3.9~pc), and a velocity
dispersion of $<30\pm5$ km/sec \citep{filippenko03}.  Based on the
velocity dispersion, \citet{filippenko03} suggest a mass for the NC of
$\lesssim6.2\times10^6$~M$\odot$.  Using the method described in
\S\ref{samplesec}.1 to derive NC masses in late-type galaxies, we
estimate the NC mass to be $1.1\times10^6$~M$_\odot$.  The AGN is one
of the nearest and least luminous Seyfert 1 nuclei
\citep{filippenko89}.  
The black hole has a measured mass of $3.6\pm1.1\times10^5$~M$\odot$
\citep{peterson05}, roughly one-third the mass of the NC.

\citet{graham07} note two elliptical/lenticular galaxies, NGC~3384 and
7457, which have measured black hole masses \citep{tremaine02} and
apparent nuclear star clusters.  Although the nuclear sources in both
cases are unresolved in HST observations presented by
\citet{ravindranath01}, both nuclear spectra are classified as
absorption spectra by HFS97, suggesting
they are in fact nuclear star clusters.  Using distances from
\citet{jensen03}, we find the $M_H$ of the NCs of $-15.44$ and $-15.71$ for
NGC~3384 and 7457 respectively, corresponding to NC mass estimates of
$2.1$ and $2.7 \times 10^7$~M$_\odot$ (\S\ref{samplesec}.1).

Four sources in \citet{scarlata04} appear to have compact nuclear star
clusters and are classified as transition or AGN objects by HFS97.
NGC~4321 (M100) and NGC~5921 are both Sbc galaxies with bright NCs
($M_R\sim-14$) classified as transition objects in HFS97.  NGC~4321 is
undetected at X-ray and radio wavelengths despite targeted
observations \citep{ho01,nagar05}.  The two other galaxies (NGC~6384
and 6951) are in our sample and are classified as LINER and Seyfert
respectively (\S\ref{agnsec}.1).  Also in our sample is the Sb galaxy
ESO205-G7, mentioned in \citet{rossa06} as having a broad
emission-line spectrum based on unpublished VLT/UVES spectra; however,
this galaxy has no available spectrum or X-ray detection.  
Finally, the recent work by \citet{gonzalezdelgado07} analyzes HST WFPC2
archival data of AGN in the HFS97 survey, and includes observations of
numerous objects that may host both NCs and AGN, including six of the
galaxies for which we find overlap.  

Possibly related to these galaxies are the recent detections of MBHs
in the centers of two massive globular clusters thought to be the NCs
of stripped dwarf galaxies \citep[e.g.,][]{meylan01,bedin04}.
Dynamical observations of G1 and $\omega$Cen suggest they host black
holes with masses $>10^4$~M$_{\odot}$
\citep{gebhardt02,gebhardt05,noyola06,rasio06}.  In G1, this claim has
been further strengthened by detection of X-ray and radio emission
consistent with a $2\times10^4$~M$_{\odot}$ black hole
\citep{pooley06,ulvestad07}.  There is also much more tentative
evidence for MBHs in other globular clusters
\citep{maccarone07,trenti06,mclaughlin06a,vandenbosch06}.

In summary, we find six additional galaxies that have both NCs and
MBHs, of which four have measured black hole masses.

\section{Discussion} \label{dissec}

\begin{figure*}
\plotone{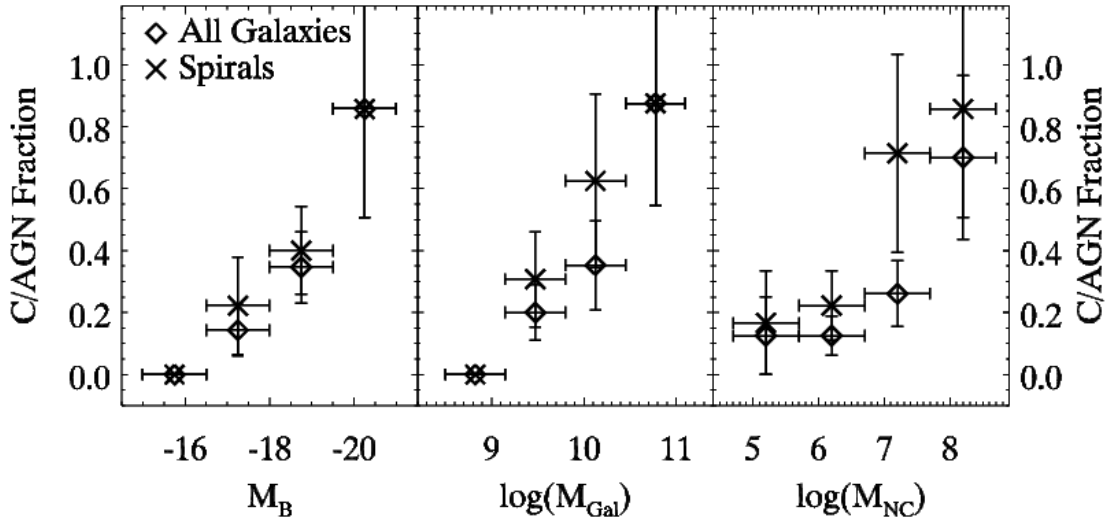}
\caption{Fraction of the 75 galaxies in the spectroscopic sample
classified as composite or AGN objects as a function of galaxy B-band
absolute magnitude {\it (left)}, galaxy mass {\it (center)}, and NC
mass {\it (right)}.  In all panels, diamonds indicate the detection
rate for all galaxies, while X's include only the spiral galaxies
(T$>$0).  The vertical bars show Poisson errors, while the horizontal
bars indicate the width of each bin.}
\label{agnfracfig}
\end{figure*}

We now discuss the most interesting results from our study.  These
include the demographics of galaxies with both AGN and NCs and the
relative masses of MBHs and NCs.  We also discuss our results in the
context of models of the formation of NCs and MBHs.

\subsection{Demographics of galaxies with AGN \& nuclear star clusters}

In this section we first discuss the range of properties spanned by
the galaxies with both NCs and AGN, and then examine the detailed
demographics of these galaxies in our spectroscopic sample.  

The left panel of Figure~\ref{galagnfig} shows the Hubble type, mass,
and absolute B-band magnitude of all galaxies presented in this paper
hosting both NCs and candidate AGN, including sources from the
literature (\S\ref{litsec}).  Galaxies with both AGN and NCs cross all
Hubble types and reach magnitudes as faint as $M_B \sim -16$ and
galaxy masses as low as $\sim10^9$~M$_\odot$.  The AGN$+$NC galaxies
are most common among early type spirals and brighter/more massive
galaxies \citep[e.g.,][]{decarli07}.  As discussed in
\S\ref{agnsec}.3, the sources most likely to be MBH$+$NC galaxies are
the 7 spectroscopic AGN candidates (squares in
Figure~\ref{galagnfig}), the four brightest localized X-ray sources
(black X's), and the six sources from the literature not contained
within our sample (circles).  Even these galaxies span a wide range of
Hubble types and luminosities/masses.  Thus the overlap of MBHs and
NCs is a phenomenon that occurs in all types of galaxies between
masses of $10^9$~M$_\odot$ and $10^{11}$~M$_\odot$.

To be more quantitative about which galaxies host AGN and which do
not, we limit ourselves to the 75 galaxies in the spectroscopic sample
for which both the detections {\it and} non-detections of AGN are
known.  The right panel of Figure~\ref{galagnfig} shows the
classification of all galaxies in the spectroscopic sample; triangles
represent galaxies with absorption spectra, crosses the star-forming
galaxies, diamonds the composite galaxies, and squares the AGN.
Seven galaxies are classified as AGN (9\% of the sample), and 18 as
composite or AGN (24\% of the sample).  Based on our discussion in
\S\ref{agnsec}.3, this suggests that $\gtrsim10$\% of NCs in our
spectroscopic sample host MBHs.

Most of the galaxies with AGN spectra (including the composite
objects) are among the brighter and more massive galaxies in our sample.
This trend is more clearly seen in Figure~\ref{agnfracfig}, which shows
the fraction of galaxies in the spectroscopic sample with C/AGN spectra
as a function of galaxy B-band absolute magnitude, galaxy mass, and NC
mass.  A strong correlation is seen in each case, with the highest
mass galaxies and NCs having AGN fractions of $\sim80$\%.
This trend of increasing AGN activity with increasing galaxy mass is
well documented \citep[e.g.,][]{ho97b,kauffmann03,decarli07}.  
The high detection rate
of AGN in galaxies with $M_B < -20$ and with galaxy mass of
$\sim10^{11}$~M$_\odot$ is consistent with the findings of these
previous surveys.  In particular, \citet{decarli07} find that almost
all Virgo cluster spiral galaxies with dynamical mass
$>10^{11}$~M$_\odot$ have AGN spectra, with little dependence on
Hubble type.  However, a direct comparison with previous studies is
complicated by the strong distance dependence of the detected AGN
fraction \citep{kauffmann03} and varying survey depths.

One way of considering whether NCs have any effect on the presence of
an AGN is to look at the fraction of AGN as a function of the galaxies'
$M_{NC}/M_{Gal}$ (see
Figure~\ref{nucmassfig}).  For galaxies with $M_{NC}/M_{Gal} > 10^{-3}$
the fraction of AGN is very similar (28\%) to those with
$M_{NC}/M_{Gal} < 10^{-3}$ (25\%).  This indicates that the trend 
observed in the right panel of Figure~\ref{agnfracfig} of 
increasing AGN fraction with increasing NC mass may just reflect the
correlation of NC mass with galaxy mass.

Some variation in the AGN fraction is also seen with Hubble type.
\citet{ho97b} find that transition and AGN spectra are found in
$\sim50\%$ of E and S0 galaxies, $70\%$ of Sa's, $50\%$ of Sb's, and $15\%$
of Sc-Sm's.  Unfortunately, direct comparison to our sample is
complicated by our inclusion of SDSS data, which differ from the HFS97
data in selection function and sensitivity.  Taking just the data from
their survey (blue and green points in Figure~\ref{galagnfig}), we
classify 1 of 4 E/S0's, 7 of 9 Sa-Sbc's, and 3 of 10 Sc-Sm's as 
C/AGN galaxies.
Despite the small number of galaxies (and slight differences in the
classifications), these detection fractions in our galaxies with NCs
are statistically consistent with what \citet{ho97b} find for the
survey as whole.

The spectroscopic sample does not uniformly cover our full NC galaxy
sample.  Good SDSS coverage of the Virgo clusters provides spectra for
a high fraction of ellipticals ($\sim75\%$), but the fraction of
spirals with spectra is much lower ($\sim30\%$) and is particularly
lacking for the fainter spirals.  This lack of spectra hinders our
ability to determine the demographic trends of AGN in our sample.

Overall, the demographic evidence suggests that galaxies with NCs have
AGN fractions consistent with the population of galaxies as a whole.
However, a larger, uniform sample of spectra for galaxies with
and without NCs is needed to test this conclusion.

\subsection{The relative mass of nuclear star clusters and massive black holes}

\begin{figure*}
\plotone{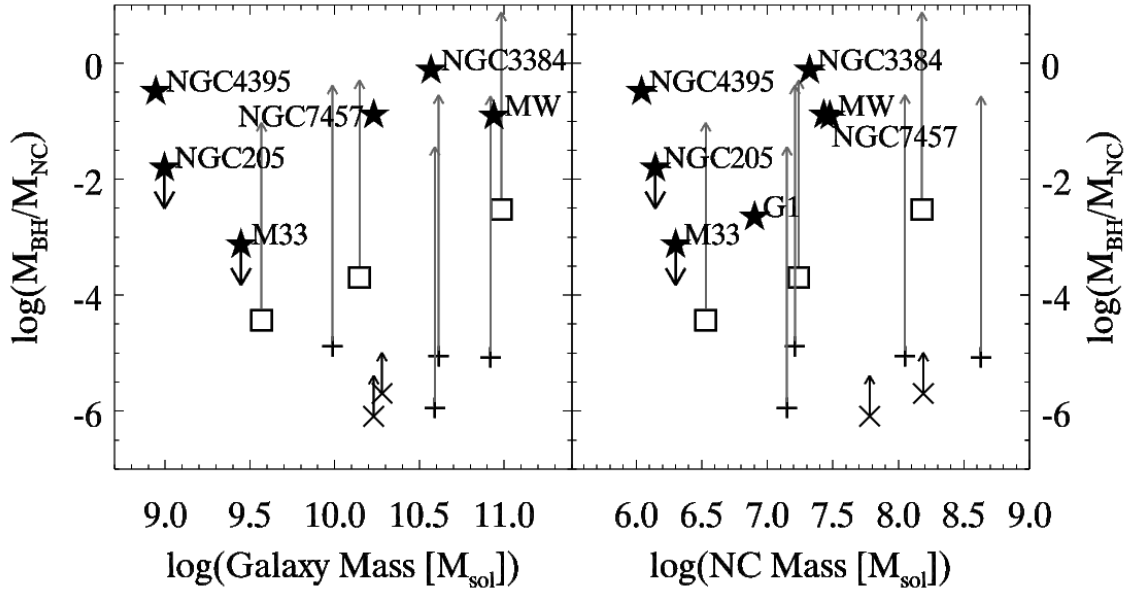}
\caption{Limits on the relative mass of nuclear star clusters and
massive black holes as a function of galaxy mass {\it (left)} and
NC mass {\it (right)}.  Stars are galaxies which have
measured black hole masses (or upper limits) in the literature
(\S3.3).  The rest of the points represent lower limits based on the
bolometric luminosity of their AGN.  Different symbols indicate
Seyferts (squares), LINERS (crosses), and X-ray sources (X's).  For
the Seyferts and LINERS, the arrow length indicates the typical
$L_{bol}/L_{Edd}$ for each class from \citet{ho04}. }
\label{cluster_bhmassfig}
\end{figure*}

In \citet{ferrarese06}, the $log(M)-\sigma$ relationship for NCs and
MBHs have similar slopes, but the normalizations differ by roughly an
order of magnitude, with NCs being more massive at a given bulge
velocity dispersion.  Similarly, \citet{rossa06} find that the
normalization of the $M_{NC}-L_{bulge,B}$ relation gives NC masses
$\sim3.3$ times more massive than the corresponding MBH relation
from \citet{marconi03}.  Based on a model for the $M_{CMO}-M_{gal}$
relation resulting from feedback of the CMO on the host galaxy,
\citet{mclaughlin06b} suggested this offset may be the result of the
reduced efficiency of NC formation feedback relative to feedback from
an accreting MBH, thus allowing the NCs to grow larger than an MBH at a
given galaxy mass.  With the present sample of NCs with AGN activity,
we have the opportunity to potentially measure the relative mass of
the NCs and MBHs {\it within the same galaxies}.

For four spiral galaxies from the literature (Milky Way, NGC~4395,
NGC~3384, NGC~7457, see \S\ref{litsec}), we can combine direct
estimates of the black hole masses and our own estimates of the
NC masses (\S\ref{samplesec}.1).
Figure~\ref{cluster_bhmassfig} shows the galaxy mass vs. $M_{BH}/M_{NC}$
ratios for these objects, which all fall between 0.1 and 1.

We can constrain the $M_{BH}/M_{NC}$ ratio from our sample galaxies
by deriving lower limits to the masses of black holes from the
bolometric luminosity of the AGN.  Both the H$\alpha$ and [OIII]
luminosities are known to be good indicators of the X-ray and
bolometric luminosities of low-luminosity AGN of all types
\citep{ho01}, including sources with significant absorption in the
X-ray \citep{panessa06}.  Because the H$\alpha$ luminosity suffers
more contamination from star-formation, we use [OIII] luminosities to
estimate L$_X$ using the \citet{panessa06} relation:
\begin{equation}
log (L_X) = 1.22\ log(L_{[OIII]}) - 7.34
\end{equation}
This yields typical $L_{[OIII]}/L_X \sim 15$ for our sample.  We note
that in the five cases in our sample where both X-ray data and C/AGN
optical spectra are available, their luminosities are consistent with
galaxies in the \citet{panessa06} sample.  In these cases we derive
the intrinsic X-ray luminosity based on the observed [OIII] luminosity
to correct for any X-ray absorption.  Then, for both our optical and
X-ray sources, we assume an $L_{bol}$/$L_{X} \sim 10$ based on the
results of the \citet{ho99} for a sample of seven low-luminosity AGN.
Note that this is a factor of $\sim$3 less than the typically assumed
$L_{bol}/L_{X}$ based on the luminosity function of luminous quasars
\citep{elvis94}.  The derived bolometric luminosities were then
divided by the Eddington luminosity for a 1~M$_\odot$ object to obtain
a lower limit on the black hole mass.  The resulting lower limits on
$M_{BH}/M_{NC}$ (actually $(L_{bol}/L_{Edd,\odot})/M_{NC}$) are shown
in Figure~\ref{cluster_bhmassfig} for the nine strongest AGN candidates
in our sample.

The highest lower limit from the sample is observed for NGC~6951,
which has $M_{BH}/M_{NC} > 3\times10^{-3}$.  The mean lower limit on
$M_{BH}/M_{NC}$ is $3\times10^{-4}$ for the three Seyfert galaxies
(squares), and $6\times10^{-6}$ for the four LINERS (crosses).  For
AGN in general (i.e., without respect to the presence of a NC), the
mean $L_{bol}/L_{Edd}$ ratios are $\sim4\times10^{-4}$ for Seyferts
and $\sim3\times10^{-5}$ for LINERS and composite objects
\citep{ho04}.  If these ratios hold for the AGN in our sample, then
our results are consistent with $M_{BH}/M_{NC}$ of near unity.  These
mean $L_{bol}/L_{Edd}$ ratios are indicated in
Figure~\ref{cluster_bhmassfig} by the length of the arrows from the
Seyfert and LINER data points.

On the other hand, a strong piece of evidence that not all NCs host
black holes of similar mass are the non-detections of black holes in
Local Group galaxies with nuclear star clusters.  In M33,
\citet{gebhardt01} place an upper limit on the black hole mass of
$1500$~M$_\odot$. Assuming a mass for the M33 NC of $\sim2\times
10^6$~M$_\odot$ \citep{kormendy93}, this non-detection implies an
$M_{BH}/M_{NC} \lesssim 10^{-3}$.  Similarly, in NGC~205, the upper
limit on the black hole mass is $2.2\times 10^4$~M$_\odot$
\citep{valluri05}, while the NC mass is $1.4\times 10^6$~M$_\odot$
\citep{derijcke06}, giving $M_{BH}/M_{NC} < 1.6 \times 10^{-2}$.

Finally, the globular cluster G1 in Andromeda, proposed to be the
stripped nuclear star cluster of a dwarf galaxy \citep{meylan01}, also
has an available measurement of $M_{BH}/M_{NC}$.  The globular cluster
mass is $8\times10^6$~M$_\odot$ \citep{baumgardt03} and the black hole
mass $1.8\times10^4$~M$_\odot$\citep{gebhardt05}, giving
$M_{BH}/M_{NC} \sim 2\times10^{-3}$.  

In summary, for four galaxies with measured black hole masses, the
ratio of $M_{BH}/M_{NC}$ ranges from 0.1 to 1.0.  
Candidate AGN in our sample also have luminosities that imply a
$M_{BH}/M_{NC}$ of near unity.  However, the black hole in the
globular cluster G1 and black hole mass upper limits in M33 and
NGC~205 show that at least some NCs have much lower $M_{BH}/M_{NC}$
or may not contain a MBH at all.  The higher $M_{BH}/M_{NC}$
measurements may result from a bias towards finding high mass black
holes in galaxies that have AGN or measurable black hole masses.
Overall, the evidence presented here suggests a wide range exists in
the relative masses of NCs and MBHs.  

\subsection{Formation Mechanisms}

Based on the scaling relations for central massive objects and the
presence of galaxies with both NCs and MBHs, there are two
possibilities for the relative formation of central massive objects.
Either the formation of MBHs could be directly linked in some way to
the formation of NCs, or NCs and MBHs could be produced by processes
which scale similarly with galaxy mass, but are otherwise unrelated.
We explore the first possibility below.  

Because NCs are the dominant objects at the centers of lower mass
galaxies \citep{ferrarese06,wehner06}, one natural possibility in hierarchical galaxy
formation scenarios is that NCs lead to the
formation of MBHs.  Formation mechanisms for NCs fall into two
classes: those that create NCs from the merging of globular clusters
due to dynamical friction \citep{tremaine75,lotz01}, and those in
which NCs are created {\it in situ} from gas accretion onto the
nucleus due to galaxy merging \citep{mihos94} or from disk gas
dynamics \citep{milosavljevic04,bekki06}.  The {\it in situ} scenario
is favored by observations that NCs in spiral galaxies have
complicated star formation histories, suggesting frequent episodic
star formation \citep{walcher06,rossa06}.  In addition, \citet{seth06}
found NCs in late-type spirals that have young stellar disks aligned
with the host galaxy disks, indicative of gas accretion onto the
nucleus.  \citet{cote06} also found support for gas accretion in the
formation of NCs in elliptical galaxies; their more massive NCs are
redder (implying higher metallicity) and more luminous than would
be expected if they formed by mergers of globular clusters.

During {\it in situ} formation, MBH formation could occur through the
merging of massive stars at the center of the cluster \citep[see
review by][]{miller04}.  For instance, dynamical modelling by
\citet{portegieszwart04} of the dense $3.5 \times 10^5$~M$_\odot$
cluster MGG~11 in M82 shows that runaway collisions of massive stars
leads to the formation of a black hole with mass $\sim0.2-1\%$ of the
mass of the cluster.  Star formation episodes in NCs could certainly
reproduce the high densities and massive star formation required for
this mode of MBH formation.  However, the MBHs in at least some of the
NCs in our sample have masses nearly equal to that of the NC;
further accretion onto the MBH would be required to explain these
cases.  Also, if MBHs were a natural consequence of NC formation, we
might expect to see a higher fraction of AGN in galaxies with NCs than
for the general galaxy population; this does not appear to be the case
(see \S\ref{dissec}.1).  These lines of evidence suggest that if there is
a link between formation of NCs and MBHs, the NCs provide at most seed
black holes which can then accrete into a more massive black hole. 

The lack of NCs in high mass galaxies also provides a clue to the
evolution of CMOs.  It is possible that during the buildup of more
massive galaxies, the NCs could be destroyed \citep[e.g., by black
hole merging;][]{milosavljevic01}, leaving behind only an MBH.
Alternatively, the presence of an MBH would likely suppress the
process of accretion and star formation in the NC due to enhanced
feedback, thus preventing further NC growth.

Theoretical work and simulations are necessary to better understand
these processes.  Specifically, further simulations of MBH
formation in massive clusters are needed that incorporate the 
episodic gas accretion thought to occur in NCs.  Modelling of NCs and MBHs in
hierarchical merging scenarios are also essential to explain the
demographics of galaxies that host NCs and/or MBHs.

\section{Conclusions and Future Work} \label{conclusions}

We have assembled a sample of 176 galaxies known to host
nuclear star clusters to study the relationship between nuclear star 
clusters and massive black holes.  We then use optical spectroscopy and radio and
X-ray data to look for AGN in this sample.  For 75 galaxies with
available optical spectra, we classify 7 galaxies as AGN and 11 as
composite objects.  X-ray catalogs from {\it Chandra, ROSAT}, and {\it
XMM} provide detections of 22 galaxies in the sample, 4 of which have
well-localized sources with X-ray luminosities indicating they are
likely AGN. Lastly, we have assembled previously published results for
$9$ galaxies that indicate they have both NCs and MBHs/AGN; three are
included in our sample.  

From this work we conclude that galaxies with
both NCs and MBHs are relatively common.  In addition, by examining 
the objects in our sample we find that:
\begin{enumerate}
\item Galaxies that host both NCs and AGN/MBHs span all Hubble
  types, have magnitudes as faint as $M_B \sim -16$, and masses as
  small as 10$^9$~M$_\odot$.  For galaxies in our sample with
  available spectra, $\gtrsim 10$\% appear to host both NCs and MBHs.
\item Galaxies with NCs have AGN detection fractions that increase
  strongly with increasing galaxy and NC mass, consistent with
  previous studies of the general galaxy population.  Variation of the
  AGN fraction with Hubble type in our sample is also consistent with
  the full Palomar survey \citep{ho97b}.  This suggests that nuclear
  clusters do not play a strong role in promoting or limiting the
  activity of massive black holes.
\item In four galaxies with NCs and measured black hole masses the
  ratios of the MBH mass to the NC mass are between 0.1 and 1.  The
  luminosity of the AGN in our sample of galaxies with NCs are also
  consistent with these mass ratios.  However, the non-detection and
  low masses of MBHs in Local Group NCs suggest a much wider range of
  MBH-to-NC mass ratios.  
\item For galaxies of the same mass, NCs in late-type
  spiral galaxies are typically an order of magnitude less massive
  than those in elliptical galaxies.  This result will be discussed
  further in a forthcoming paper.
\end{enumerate}

Although the data presented here provide compelling evidence for a
significant overlap between the MBH and NC population, the evidence
for the presence of an active MBH in many of the individual objects is quite weak.  
A number of observations could strengthen this evidence, including
high-resolution {\it Chandra} observations or {\it Spitzer} mid-IR
spectroscopy \citep[e.g.,][]{satyapal07}.  Furthermore, less than half
of the galaxies in our sample have available optical spectroscopy -- obtaining
spectroscopy for more is desirable and will help fill gaps in
parameter space (e.g., the lack of spectra of low-mass early type
spiral).  Finally, detailed studies of objects with both NCs and MBHs
will be required to improve our understanding of the connections
between these central massive objects.

Acknowledgments: The authors thank Christy Tremonti for providing
significant assistance with the SDSS spectra.  We also acknowledge
valuable advice received from Margaret Geller, Julianne
Dalcanton, David Helfand, Beth Willman, Andrew West, Nelson Caldwell,
Andr\'es Jord\'an, Bob Becker, Ramesh Narayan, Martin Elvis, and
Giuseppina Fabbiano.  Anil Seth gratefully acknowledges the support of
the CfA Postdoctoral Fellowship.  Marcel Ag\"ueros is supported by an
NSF Astronomy and Astrophysics Postdoctoral Fellowship under award
AST-0602099.

This work would not have been possible without use of the Sloan
Digital Sky Survey ({\tt http://www.sdss.org/}), the NASA
Extragalactic Database ({\tt http://nedwww.ipac.caltech.edu/}),
HyperLeda ({\tt http://leda.univ-lyon1.fr/}), and NASA's High Energy
Astrophysics Science Archive Research
Center ({\tt http://heasarc.gsfc.nasa.gov/}).

\begin{deluxetable}{lcccccccc}
\tablecaption{Optical Spectroscopy for Star-forming Galaxies \label{hiitab}}
\tablehead{
     \colhead{Galaxy}  &
     \colhead{log([OIII]/H$\beta$)}  &
     \colhead{log([NII]/H$\alpha$)}  &
     \colhead{log([SII]/H$\alpha$)}  &
     \colhead{log([OI]/H$\alpha$)}  &
     \colhead{H$\alpha$/H$\beta$\tablenotemark{a}}  &
     \colhead{SFR [M$_\odot$/yr]}  &
     \colhead{SDSS Class\tablenotemark{b}}  &
     \colhead{HFS97 Class\tablenotemark{c}} 
}

\startdata
      NGC450 & -0.03$\pm$0.04 &  -0.56$\pm$0.03 &  -0.40$\pm$0.02 &  -1.33$\pm$0.07 &      4.1 &  7.7E-03 &      HII &  \nodata \\
     UGC4499 &  0.11$\pm$0.04 &  -0.89$\pm$0.06 &  -0.40$\pm$0.03 &  -1.34$\pm$0.10 &      3.2 &  4.6E-04 &      HII &  \nodata \\
     NGC2964 & -0.55$\pm$0.10 &  -0.31$\pm$0.10 &  -0.58$\pm$0.10 &  -1.77$\pm$0.10 &      5.6 &  3.5E-01 &  \nodata &      HII \\
     NGC3346 & -0.56$\pm$0.05 &  -0.45$\pm$0.02 &  -0.50$\pm$0.02 &  -1.53$\pm$0.10 &      3.7 &  3.4E-03 &      HII &      HII \\
     NGC3913 & -0.51$\pm$0.08 &  -0.40$\pm$0.03 &  -0.45$\pm$0.03 &  -1.35$\pm$0.12 &      3.0 &  1.2E-03 &      HII &  \nodata \\
     NGC3949 & -0.66$\pm$0.10 &  -0.40$\pm$0.10 &  -0.36$\pm$0.10 &  -1.30$\pm$0.30 &  \nodata &  \nodata &  \nodata &      HII \\
    A1156+52 & -0.13$\pm$0.06 &  -0.63$\pm$0.04 &  -0.39$\pm$0.03 &  -1.77$\pm$0.30 &      3.8 &  2.0E-03 &      HII &  \nodata \\
     NGC4144 &  0.08$\pm$0.04 &  -0.87$\pm$0.06 &  -0.31$\pm$0.03 &  -1.20$\pm$0.10 &      3.2 &  2.0E-04 &      HII &      HII \\
     NGC4183 & -0.08$\pm$0.03 &  -0.62$\pm$0.02 &  -0.45$\pm$0.02 &  -1.58$\pm$0.07 &      5.5 &  1.1E-02 &      HII &      HII \\
     NGC4244 & -0.22$\pm$0.11 &  -0.70$\pm$0.09 &  -0.43$\pm$0.06 &  -1.32$\pm$0.24 &      3.9 &  1.6E-04 &      HII &      HII \\
     NGC4384 & -0.18$\pm$0.02 &  -0.53$\pm$0.01 &  -0.56$\pm$0.01 &  -1.67$\pm$0.05 &      3.2 &  8.3E-02 &      HII &  \nodata \\
     VCC1250 & -0.29$\pm$0.23 &  -0.34$\pm$0.12 &  -0.45$\pm$0.16 &  -0.92$\pm$0.25 &      3.9 &  1.8E-03 &      HII &  \nodata \\
     NGC4618 &  0.14$\pm$0.01 &  -0.71$\pm$0.01 &  -0.54$\pm$0.01 &  -1.62$\pm$0.04 &      3.1 &  3.7E-03 &      HII &      HII \\
     NGC5585 &  0.23$\pm$0.10 &  -0.82$\pm$0.10 &  -0.50$\pm$0.10 &  -1.77$\pm$0.10 &  \nodata &  \nodata &  \nodata &      HII \\
     NGC5584 & -0.50$\pm$0.04 &  -0.50$\pm$0.02 &  -0.53$\pm$0.02 &  -1.65$\pm$0.10 &      3.5 &  7.7E-03 &      HII &  \nodata \\
     NGC5669 & -0.40$\pm$0.02 &  -0.57$\pm$0.01 &  -0.50$\pm$0.01 &  -1.82$\pm$0.07 &      3.5 &  1.6E-02 &      HII &      HII \\
     NGC5774 & -0.41$\pm$0.17 &  -0.41$\pm$0.05 &  -0.22$\pm$0.04 &  -1.14$\pm$0.14 &      4.2 &  2.2E-03 &      HII &  \nodata \\
     NGC6239 &  0.23$\pm$0.01 &  -0.76$\pm$0.01 &  -0.62$\pm$0.01 &  -1.86$\pm$0.03 &      4.5 &  6.5E-02 &      HII &  \nodata \\
\enddata
\tablenotetext{a}{Galaxies with no data for their
H$\alpha$/H$\beta$ ratio and SFR are those for which the HFS97 data were not
photometric.}  
\tablenotetext{b}{The classification of HII indicates a star-forming spectra (\S\ref{agnsec}.1).}
\tablenotetext{c}{This column gives {\it our} classification of the HFS97 line ratios.}
\end{deluxetable}

\begin{deluxetable}{lcccccccc}
\tablecaption{Optical Spectroscopy for Composite and AGN Galaxies \label{agntab}}
\tablehead{
     \colhead{Galaxy}  &
     \colhead{log([OIII]/H$\beta$)}  &
     \colhead{log([NII]/H$\alpha$)}  &
     \colhead{log([SII]/H$\alpha$)}  &
     \colhead{log([OI]/H$\alpha$)}  &
     \colhead{H$\alpha$/H$\beta$ \tablenotemark{a}}  &
     \colhead{$L_{[OIII]}$/$L_\odot$}  &
     \colhead{SDSS Class\tablenotemark{b}}  &
     \colhead{HFS97 Class\tablenotemark{c}} 
}

\startdata
      NGC428 & -0.12$\pm$0.10 &  -0.25$\pm$0.10 &   0.12$\pm$0.10 &        $<$-0.72 &  \nodata &  \nodata &  \nodata &     C/L2 \\
     NGC1042 & -0.01$\pm$0.02 &  -0.26$\pm$0.01 &  -0.45$\pm$0.02 &  -1.20$\pm$0.04 &      3.6 &  6.7E+04 &        C &  \nodata \\
     NGC3177 & -0.43$\pm$0.03 &  -0.20$\pm$0.01 &  -0.58$\pm$0.02 &  -1.40$\pm$0.05 &      5.6 &  3.7E+05 &        C &  \nodata \\
     NGC3259 &  0.72$\pm$0.04 &  -0.13$\pm$0.03 &  -0.41$\pm$0.04 &  -0.94$\pm$0.06 &      4.9 &  5.6E+05 &   AGN/S2 &  \nodata \\
     NGC3423 & -0.38$\pm$0.10 &  -0.31$\pm$0.05 &  -0.38$\pm$0.05 &  -1.15$\pm$0.14 &      3.4 &  3.4E+03 &        C &        C \\
     NGC3928 & -0.25$\pm$0.02 &  -0.31$\pm$0.01 &  -0.51$\pm$0.01 &  -1.47$\pm$0.04 &      3.2 &  1.3E+05 &        C &  \nodata \\
     NGC4206 & -0.13$\pm$0.15 &  -0.31$\pm$0.09 &  -0.26$\pm$0.08 &  -1.17$\pm$0.30 &      3.9 &  2.0E+03 &        C &  \nodata \\
    NGC4411B &  0.39$\pm$0.37 &  -0.19$\pm$0.10 &  -0.27$\pm$0.12 &  -0.91$\pm$0.25 &      7.2 &  3.7E+04 &   AGN/S2 &  \nodata \\
     NGC4517 &  0.09$\pm$0.17 &  -0.29$\pm$0.10 &  -0.25$\pm$0.10 &  -1.27$\pm$0.10 &      7.7 &  5.2E+04 &  \nodata &     C/L2 \\
     VCC1619 &  0.27$\pm$0.10 &  -0.26$\pm$0.10 &   0.05$\pm$0.10 &  -0.77$\pm$0.10 &  \nodata &  \nodata &  \nodata &     C/L2 \\
     NGC4625 & -0.35$\pm$0.10 &  -0.31$\pm$0.05 &  -0.33$\pm$0.05 &  -1.00$\pm$0.10 &      3.2 &  2.1E+03 &        C &  \nodata \\
     NGC4750 &  0.24$\pm$0.10 &   0.46$\pm$0.10 &   0.30$\pm$0.10 &  -0.22$\pm$0.10 &      3.3 &  2.0E+05 &  \nodata &   AGN/L1.9 \\
     NGC5377 &  0.30$\pm$0.17 &   0.33$\pm$0.10 &   0.22$\pm$0.10 &  -0.60$\pm$0.10 &      4.6 &  5.8E+05 &  \nodata &   AGN/L2 \\
     NGC5678 & -0.17$\pm$0.08 &  -0.22$\pm$0.03 &  -0.48$\pm$0.05 &  -1.20$\pm$0.12 &      7.0 &  3.7E+05 &        C &        C \\
     NGC5806 & -0.05$\pm$0.07 &  -0.15$\pm$0.05 &  -0.38$\pm$0.06 &  -1.07$\pm$0.14 &      4.1 &  8.4E+04 &        C &        C \\
     NGC5879 &  0.08$\pm$0.06 &  -0.05$\pm$0.05 &  -0.05$\pm$0.05 &  -0.68$\pm$0.08 &      3.6 &  2.6E+04 &   AGN/L2 &      AGN \\
     NGC6384 &  0.18$\pm$0.17 &   0.24$\pm$0.10 &   0.03$\pm$0.10 &        $<$-0.82 &      3.1 &  6.8E+03 &  \nodata &   AGN/L2 \\
     NGC6951 &  0.82$\pm$0.10 &   0.39$\pm$0.10 &  -0.04$\pm$0.10 &  -0.64$\pm$0.10 &     16.7 &  3.0E+07 &  \nodata &   AGN/S2 \\
\enddata
\tablenotetext{a}{Galaxies with no data for their H$\alpha$/H$\beta$
ratio and [OIII] luminosity are those for which the HFS97 data were
not photometric.}  
\tablenotetext{b}{Classes are defined in \S\ref{agnsec}.1, ``C''=composite, ``L''=LINER and ``S''=Seyfert.}

\tablenotetext{c}{This column gives {\it our} classification of the HFS97 line ratios.}
\end{deluxetable}

\begin{deluxetable}{lccccccc}
\tablewidth{0pt} \tabletypesize{\scriptsize}
\setlength{\tabcolsep}{0.03in} 
\tablecaption{Radio data for the $13$ galaxies with FIRST matches.\label{radio_data}}
\tablehead{ 
\colhead{}     & \colhead{Optical} & \colhead{Separation}    & \colhead{Integrated} & \colhead{RMS}  & \colhead{Major }    & \colhead{Minor}     & \colhead{$L_{1.4 GHz}$} \\
\colhead{Galaxy} & \colhead{Class}   & \colhead{NC (\asec)}  & \colhead{Flux (mJy)} & \colhead{(\asec)} & \colhead{Axis (\asec)} & \colhead{Axis (\asec)} & \colhead{(W Hz$^{-1}$)} 
}
\startdata
NGC 2964 & HII     & $0.2$ & $28.38$ & $0.14$ & $6.98$ & $4.99$ & $1.5\times 10^{21}$ \\
NGC 3177 & C       & $1.0$   & $26.21$ & $0.24$ & $9.47$ & $7.56$ & $1.3\times 10^{21}$ \\
NGC 3277 & \nodata & $0.6$ & $3.29$ & $0.14$ & $7.40$ & $6.77$ & $1.9\times 10^{20}$ \\
NGC 3928 & C       & $0.7$ & $8.55$ & $0.14$ & $10.11$ & $8.99$ & $2.6\times 10^{20}$ \\
NGC 3949 & HII     & $24.9$ & $21.08$ & $0.14$ & $23.21$ & $17.68$ & $5.4\times 10^{20}$ \\
NGC 4030 & \nodata & $10.0$ & $34.26$ & $0.16$ & $41.07$ & $26.28$ & $1.8\times 10^{21}$ \\
NGC 4144 & HII     & $14.9$ & $1.56$ & $0.14$ & $3.82$ & $0.00$ & $9.8\times 10^{18}$ \\
VCC 437 & \nodata  & $25.9$ & $4.94$ & $0.13$ & $9.19$ & $8.38$ & $1.6\times 10^{20}$ \\
NGC 4384 & HII     & $3.3$ & $12.52$ & $0.15$ & $14.39$ & $10.39$ & $2.3\times 10^{21}$ \\
NGC 4701 & \nodata & $3.6$ & $5.93$ & $0.15$ & $18.66$ & $8.88$ & $8.8\times 10^{19}$ \\
NGC 5377 & AGN/L2 & $1.4$ & $3.26$ & $0.14$ & $6.94$ & $6.61$ & $3.8\times 10^{20}$ \\
NGC 5678\tablenotemark{a} & C & $7.2$ & $68.39$ & $0.14$ & $44.03$ & $32.26$ & $6.6\times 10^{21}$ \\
NGC 6239\tablenotemark{a} & HII & $0.8$ & $4.11$ & $0.17$ & $11.96$ & $5.43$ & $1.4\times 10^{20}$ \\
\enddata
\tablenotetext{a}{Flagged as a possible sidelobe of a nearby bright source. NGC 5377 has two other fainter FIRST sources within $20\asec$ of the NC position; NGC 6239 has no other FIRST sources within $60\asec$. It is unclear why these radio sources are flagged.}
\tablecomments{All of these galaxies were also detected by NVSS.}
\end{deluxetable}

\begin{deluxetable}{lclccc}
\tablewidth{0pt} \tabletypesize{\scriptsize}
\setlength{\tabcolsep}{0.03in} 
\tablecaption{X-ray data for the $22$ galaxies with {\it Chandra}, {\it ROSAT}, or {\it XMM} matches.\label{xray_data}}
\tablehead{ 
\colhead{}     & \colhead{Optical\tablenotemark{a}} & \colhead{X-ray}       & \colhead{Separation} & \colhead{Extent}    & \colhead{$L_X$\tablenotemark{b}, $2-10$ keV} \\
\colhead{Galaxy} & \colhead{Class} & \colhead{Source Name} & \colhead{NC (\asec)}       & \colhead{(\asec)} & \colhead{(erg s$^{-1}$)}
}
\startdata
\multicolumn{5}{c}{{\it Chandra}}\\
\tableline
VCC 751  & \nodata  & X122448.46$+$181141.8   & $1.3$ & $2.0\times2.0$ & $1.58\times10^{38}$\\
VCC 828  & ABS      & X122541.63$+$124837.3   & $1.3$ & $5.0\times5.0$ & $3.25\times10^{38}$\\
VCC 1192 & \nodata  & X122930.21$+$075934.6   & $0.8$ & $0.7\times0.6$ & $2.37\times10^{38}$\\
NGC 4750 & AGN/L1.9   & X125007.34$+$725228.8   & $1.0$ & $0.4\times0.4$ & $1.05\times10^{40}$ \\
NGC 5678 & C        & X143205.54$+$575517.2   & $0.5$ & $0.7\times0.3$ & $6.67\times10^{38}$\\
NGC 5774 & HII      & X145342.77$+$033503.2   & $4.9$ & $0.7\times0.6$ & $1.17\times10^{38}$\\
NGC 5879 & AGN/L2   & X150946.72$+$570000.2   & $1.0$ & $1.0\times0.5$ & $2.51\times10^{38}$\\
\tableline
\multicolumn{5}{c}{{\it ROSAT}}\\
\tableline
NGC 600  & \nodata & 2RXP J013305.7$-$071835 & $9.8$  & $0$     & $8.10\times10^{39}$\\
NGC 1385\tablenotemark{c} & \nodata & 1BMW 033728.0$-$243003  & $4.1$  & $14$    & $1.13\times10^{40}$\\
NGC 2566\tablenotemark{d} & \nodata & 1RXS J081847.0$-$252922 & $40.7$ & $0$  & $2.69\times10^{40}$\\
NGC 3259 & AGN/S2  & RX J103234$+$65024      & $9.4$  & $0$     & $1.27\times10^{40}$\\
NGC 3445 & \nodata & 1WGA J1054.5$+$5659     & $9.4$  & \nodata & $2.66\times10^{39}$\\
NGC 4030 & \nodata & RX J120023$-$01055      & $10.8$ & $9$     & $1.47\times10^{40}$\\
NGC 4540\tablenotemark{d} & \nodata & 1RXS J123456.0$+$153314 & $74.1$ & $0$  & $7.60\times10^{40}$\\
NGC 6000\tablenotemark{c} & \nodata & 1RXH J154949.7$-$292310 & $3.6$ & $0$ & $5.83\times10^{40}$\\
\tableline
\multicolumn{5}{c}{{\it XMM}}\\
\tableline
NGC 1493 & \nodata & 2XMM J035727.3$-$461239 & $4.5$ & $<6$ & $3.51\times10^{38}$ \\
VCC 1250 & HII     & 2XMM J122959.1$+$122052 & $3.0$ & $<6$ & $3.23\times10^{38}$ \\
VCC 1283 & ABS     & 2XMM J123018.3$+$133440 & $0.7$ & $<6$ & $2.82\times10^{38}$ \\
VCC 1355 & \nodata & 2XMM J123120.2$+$140656 & $1.8$ & $<6$ & $4.37\times10^{38}$ \\
NGC 4517 & C       & 2XMM J123245.4$+$000655 & $1.8$ & $<6$ & $7.25\times10^{38}$ \\
NGC 6951 & AGN/S2  & 2XMM J203714.0$+$660619 & $0.5$ & $<6$ & $5.39\times10^{39}$ \\
NGC 7418 & \nodata & 2XMM J225636.0$-$370145 & $1.5$ & $<6$ & $6.37\times10^{39}$ \\
\enddata
\tablenotetext{a}{Classes identified in \S\ref{agnsec}.1; ``ABS''=absorption spectrum.}
\tablenotetext{b}{Assuming $\Gamma = 1.8$.}
\tablenotetext{c}{{\it ROSAT}/HRI detection.}
\tablenotetext{d}{Tentatively associated with X-ray source.}
\end{deluxetable}

\clearpage
\LongTables
\begin{deluxetable}{lccccccccc}
\tablecaption{Galaxies With Nuclear Star Clusters Sample \label{sampletab}}
\tablehead{
     \colhead{Galaxy}  &
     \colhead{Type}  &
     \colhead{T}  &
     \colhead{D [Mpc]}  &
     \colhead{$M_B$}  &
     \colhead{log($M_{gal}/M_\odot$)}  &
     \colhead{Source\tablenotemark{a}}  &
     \colhead{NC $r_{eff}$\tablenotemark{b}}  &
     \colhead{log($M_{NC}/M_\odot$)}  &
     \colhead{Spectra\tablenotemark{c}} 
} 

\startdata
        NGC275 &  SBc &  6.0 & 24.0 & -19.1 & \nodata & B02 & 0.05 & 6.3 &  None \\
        NGC289 & SBbc &  3.9 & 27.4 & -20.5 &    10.9 & C02 & 0.13 & 7.9 &  None \\
        NGC300 &  Scd &  6.9 &  1.9 & -17.6 &     9.3 & B02 & 0.13 & 6.0 &  None \\
       NGC337a & SABd &  7.9 & 14.3 & -17.9 &     9.4 & B02 & 0.06 & 5.3 &  None \\
        NGC406 &   Sc &  4.9 & 17.5 & -18.2 &     9.7 & C02 & 0.10 & 5.8 &  None \\
        NGC428 & SABm &  8.6 & 15.9 & -19.1 &     9.8 & B02 & 0.05 & 6.5 & HFS97 \\
        NGC450 & SABc &  5.9 & 24.4 & -19.2 &     9.7 & B02 & 0.11 & 6.1 &  SDSS \\
        NGC600 &  Scd &  6.9 & 25.1 & -19.0 &     9.9 & B02 & 0.06 & 6.2 &  None \\
        NGC853 &   Sm &  8.7 & 21.0 & -15.9 &     9.6 & B02 & 0.05 & 6.1 &  None \\
        NGC986 &  Sab &  2.3 & 24.9 & -20.3 &    10.9 & C02 & 0.11 & 7.9 &  None \\
       NGC1042 & SABc &  6.0 & 18.0 & -19.6 &    10.0 & B02 & 0.05 & 6.5 &  SDSS \\
       NGC1325 & SBbc &  4.0 & 20.1 & -19.3 &    10.3 & C98 & 0.09 & 7.1 &  None \\
    ESO548-G10 &  Scd &  6.8 & 16.5 & -17.7 &     9.4 & C97 & 0.07 & 6.9 &  None \\
      ESO358-5 & SABm &  8.8 & 20.0 & -16.7 &     9.4 & B02 & 0.05 & 5.9 &  None \\
       NGC1345 &   Sc &  5.1 & 19.5 & -17.1 &     9.2 & C02 & 0.05 & 6.1 &  None \\
    ESO548-G29 & SABb &  3.4 & 16.2 & -16.7 & \nodata & C02 & 0.04 & 5.9 &  None \\
      ESO418-8 & SABd &  7.9 & 14.1 & -16.7 &     9.4 & B02 & 0.05 & 5.5 &  None \\
       NGC1385 &   Sc &  5.9 & 18.6 & -19.9 &    10.3 & C98 & 0.04 & 6.4 &  None \\
    ESO482-G17 &  Sab &  2.2 & 18.5 & -16.1 &    10.0 & C97 & 0.15 & 6.7 &  None \\
    ESO549-G18 & SABc &  5.0 & 20.0 & -17.9 &     9.9 & C02 & 0.17 & 6.2 &  None \\
       NGC1483 & SBbc &  4.0 & 12.6 & -17.3 &     9.1 & C02 & 0.03 & 5.3 &  None \\
       NGC1493 &  SBc &  6.0 & 11.3 & -18.4 &     9.6 & B02 & 0.06 & 6.4 &  None \\
     ESO202-41 &  SBm &  9.0 & 19.8 & -16.3 &     9.3 & B02 & 0.06 & 5.0 &  None \\
       NGC1688 &  SBc &  6.2 & 13.4 & -18.5 &     9.5 & C02 & 0.06 & 6.3 &  None \\
       NGC1892 &   Sc &  5.8 & 15.2 & -18.1 &     9.6 & C02 & 0.05 & 6.7 &  None \\
       NGC2082 &  SBb &  3.2 & 15.3 & -18.2 &     9.8 & C02 & 0.06 & 6.3 &  None \\
       NGC2104 &  SBm &  8.9 & 12.8 & -17.3 &     9.1 & C97 & 0.13 & 5.7 &  None \\
     ESO205-G7 &  SBb &  3.0 & 24.7 & -16.6 &     9.7 & C02 & 0.12 & 7.6 &  None \\
       NGC2139 &   Sc &  5.9 & 23.6 & -19.9 &    10.1 & B02 & 0.07 & 5.9 &  None \\
       UGC3574 &   Sc &  5.9 & 23.3 & -17.8 & \nodata & B02 & 0.06 & 6.1 &  None \\
       NGC2397 &  SBb &  3.1 & 15.1 & -18.2 &    10.1 & C97 & 0.27 & 8.5 &  None \\
       UGC3826 & SABc &  6.5 & 27.8 & -17.6 & \nodata & B02 & 0.07 & 5.6 &  None \\
       NGC2566 &   Sb &  2.8 & 21.0 & -19.7 &    10.7 & C02 & 0.11 & 8.7 &  None \\
       NGC2552 & SABm &  9.0 & 10.0 & -17.3 &     9.0 & B02 & 0.05 & 5.8 &  SDSS \\
       UGC4499 &   Sd &  7.9 & 12.6 & -16.2 &     8.7 & B02 & 0.07 & 4.9 &  SDSS \\
       NGC2758 & SBbc &  3.9 & 26.1 & -17.9 &    10.0 & C02 & 0.07 & 6.5 &  None \\
       NGC2763 &  SBc &  5.6 & 25.4 & -19.4 &    10.1 & B02 & 0.07 & 5.9 &  None \\
       NGC2805 & SABc &  6.9 & 28.2 & -20.5 &    10.0 & B02 & 0.06 & 6.7 &  None \\
       UGC4988 &   Sm &  8.7 & 24.3 & -16.6 &     9.1 & B02 & 0.05 & 5.8 &  SDSS \\
     ESO498-G5 & SABb &  4.3 & 32.4 & -18.5 &    10.1 & C02 & 0.05 & 7.3 &  None \\
       UGC5015 & SABd &  7.7 & 25.4 & -16.6 &     9.1 & B02 & 0.06 & 5.9 &  SDSS \\
IRAS09312-3248 &   Sc &  6.0 & 10.6 & -17.4 &     9.6 & S06 &    * & 5.0 &  None \\
       NGC2964 &  Sbc &  4.1 & 20.6 & -19.5 &    10.5 & C02 & 0.14 & 7.8 & HFS97 \\
       NGC3045 &   Sb &  3.0 & 30.5 & -18.5 &     9.9 & C98 & 0.08 & 6.8 &  None \\
    ESO499-G37 & SABc &  6.2 & 11.5 & -16.8 &     8.4 & C02 & 0.28 & 5.8 &  None \\
       NGC3177 &   Sb &  3.0 & 20.1 & -18.5 &    10.1 & C98 & 0.09 & 8.1 &  SDSS \\
       NGC3259 & SABb &  3.6 & 27.5 & -19.3 &    10.1 & C02 & 0.11 & 7.2 &  SDSS \\
       NGC3277 &  Sab &  1.8 & 22.0 & -19.2 &    10.4 & C02 & 0.15 & 8.3 &  None \\
       NGC3346 &  SBc &  6.0 & 18.9 & -19.0 & \nodata & B02 & 0.04 & 6.1 &  S\&H \\
       NGC3423 &   Sc &  6.0 & 14.7 & -19.2 &     9.9 & B02 & 0.06 & 6.5 &  S\&H \\
       NGC3455 & SABb &  3.1 & 16.9 & -16.8 &     9.5 & C02 & 0.04 & 6.8 &  None \\
       NGC3445 & SABm &  8.9 & 32.1 & -19.6 &     9.9 & B02 & 0.05 & 6.7 &  None \\
       NGC3501 &   Sc &  5.9 & 17.4 & -17.7 & \nodata & S06 &    * & 5.9 &  None \\
       NGC3782 &  Scd &  6.6 & 13.6 & -17.6 & \nodata & B02 & 0.05 & 5.4 &  None \\
       NGC3885 & S0-a &  0.2 & 24.1 & -19.0 &     9.9 & C97 & 0.22 & 8.4 &  None \\
       NGC3906 & SBcd &  6.8 & 16.9 & -17.6 &     9.3 & B02 & 0.06 & 5.4 &  None \\
       NGC3913 &  Scd &  6.6 & 17.1 & -17.8 &     9.4 & B02 & 0.25 & 5.3 &  SDSS \\
       NGC3928 &    E & -4.5 & 15.8 & -17.8 &     9.7 & C02 & 0.10 & 7.4 &  SDSS \\
       NGC3949 &  Sbc &  4.0 & 14.6 & -19.5 &     9.9 & C02 & 0.05 & 6.9 & HFS97 \\
    ESO572-G22 & SBcd &  6.7 & 26.3 & -17.1 &     9.6 & C02 & 0.04 & 6.2 &  None \\
     ESO504-30 &  SBd &  7.7 & 24.0 & -17.0 &     8.9 & B02 & 0.06 & 5.9 &  None \\
       UGC6931 &  SBm &  9.0 & 20.6 & -16.0 & \nodata & B02 & 0.05 & 5.3 &  None \\
      A1156+52 &  SBc &  5.8 & 18.8 & -18.1 &     9.4 & B02 & 0.05 & 5.7 &  SDSS \\
       NGC4027 &  SBd &  7.8 & 22.8 & -20.2 &    10.5 & B02 & 0.07 & 5.9 &  None \\
       NGC4030 &  Sbc &  4.0 & 21.1 & -20.4 &    10.9 & C98 & 0.09 & 8.0 &  None \\
       NGC4144 & SABc &  6.0 &  7.2 & -17.2 &     9.0 & S06 & 0.05 & 4.8 &  S\&H \\
         VCC33 &   E? & -2.4 & 16.5 & -16.3 & \nodata & C06 & 0.03 & 5.7 &  SDSS \\
       NGC4183 &   Sc &  5.9 & 16.2 & -18.2 &     9.6 & S06 &    * & 5.9 &  S\&H \\
        VCC140 & S0-a &  0.3 & 16.5 & -16.7 & \nodata & C06 & 0.03 & 5.7 &  SDSS \\
       NGC4204 &  SBd &  7.9 & 13.9 & -16.9 & \nodata & B02 & 0.07 & 5.4 &  None \\
       NGC4206 &  Sbc &  4.0 & 11.3 & -17.4 &     9.5 & S06 & 0.18 & 6.8 &  SDSS \\
        VCC200 &    E & -4.1 & 16.5 & -16.3 &     9.0 & C06 & 0.05 & 5.4 &  SDSS \\
        VCC230 &   E? & -2.5 & 16.5 & -15.8 &     8.8 & C06 & 0.04 & 6.5 &  SDSS \\
       NGC4244 &   Sc &  6.1 &  4.3 & -17.7 &     9.3 & S06 & 0.16 & 6.5 &  S\&H \\
        VCC437 &   S0 & -2.5 & 16.5 & -16.5 &     9.3 & C06 & 0.09 & 6.6 &  None \\
       NGC4299 & SABd &  8.3 & 16.8 & -18.4 &     9.3 & B02 & 0.05 & 6.0 &  None \\
        VCC538 & E-SO & -3.3 & 16.5 & -15.6 &     8.8 & C06 & 0.03 & 6.2 &  None \\
        VCC543 & S0-a & -1.0 & 16.5 & -16.3 & \nodata & C06 & 0.16 & 5.8 &  SDSS \\
        VCC698 &   S0 & -2.0 & 16.5 & -17.5 &     9.8 & C06 & 0.04 & 6.8 &  SDSS \\
        VCC751 & S0-a &  0.1 & 16.5 & -15.7 &     9.2 & C06 & 0.05 & 6.1 &  None \\
       NGC4384 &   Sa &  1.0 & 39.3 & -19.4 &    10.1 & C02 & 0.06 & 6.3 &  SDSS \\
        VCC784 & E-SO & -2.8 & 16.5 & -18.4 &    10.1 & C06 & 0.16 & 7.7 & HFS97 \\
        VCC828 &    E & -4.8 & 16.5 & -18.1 &    10.0 & C06 & 0.21 & 7.5 &  SDSS \\
        VCC856 & E-SO & -2.9 & 16.5 & -16.6 &     9.3 & C06 & 0.16 & 7.1 &  SDSS \\
       NGC4416 &   Sc &  5.9 & 20.8 & -18.5 & \nodata & B02 & 0.22 & 4.9 &  None \\
      NGC4411B & SABc &  6.1 & 19.1 & -18.4 &     9.6 & B02 & 0.06 & 6.5 &  SDSS \\
       VCC1075 &   E? & -2.7 & 16.5 & -16.0 &     8.9 & C06 & 0.04 & 6.1 &  SDSS \\
       VCC1087 &    E & -4.1 & 16.5 & -16.7 &     9.4 & C06 & 0.03 & 6.7 &  SDSS \\
       VCC1125 &   S0 & -1.8 & 16.5 & -18.2 &    10.0 & C06 & 0.06 & 6.4 &  SDSS \\
       VCC1146 &    E & -4.8 & 16.5 & -18.2 &    10.0 & C06 & 0.78 & 8.7 &  None \\
       VCC1185 &    E & -4.2 & 16.5 & -15.6 &     8.8 & C06 & 0.06 & 6.2 &  SDSS \\
       VCC1192 &    E & -4.8 & 16.5 & -16.2 &     9.4 & C06 & 0.12 & 7.1 &  None \\
       VCC1199 &    E & -4.1 & 16.5 & -15.5 &     8.9 & C06 & 0.08 & 6.9 &  None \\
       VCC1242 &   S0 & -2.0 & 16.5 & -18.6 &    10.2 & C06 & 0.04 & 7.1 &  None \\
       VCC1250 & E-SO & -3.0 & 16.5 & -17.9 &     9.9 & C06 & 0.03 & 7.0 &  SDSS \\
       VCC1261 &    E & -4.8 & 16.5 & -17.4 &     9.5 & C06 & 0.04 & 7.0 &  SDSS \\
       VCC1283 &   S0 & -1.8 & 16.5 & -17.6 &     9.9 & C06 & 0.05 & 6.7 &  SDSS \\
       NGC4487 &   Sc &  5.9 & 14.7 & -19.0 &     9.8 & B02 & 0.05 & 6.6 &  None \\
       VCC1355 &    E & -4.8 & 16.5 & -16.5 & \nodata & C06 & 0.04 & 6.2 &  None \\
      NGC4496A &  SBd &  7.5 & 15.0 & -18.8 &     9.7 & B02 & 0.05 & 5.7 &  None \\
       VCC1407 &    E & -4.1 & 16.5 & -16.0 &     9.0 & C06 & 0.14 & 6.4 &  SDSS \\
       VCC1422 &    E & -4.8 & 16.5 & -17.1 &     9.4 & C06 & 0.04 & 6.7 &  SDSS \\
       VCC1431 &    E & -4.0 & 16.5 & -16.7 &     9.4 & C06 & 0.24 & 6.8 &  SDSS \\
       VCC1440 &    E & -4.4 & 16.5 & -16.1 &     9.2 & C06 & 0.06 & 6.9 &  SDSS \\
       NGC4517 &   Sc &  6.0 & 16.5 & -20.0 &    10.6 & S06 & 0.04 & 6.8 & HFS97 \\
       VCC1488 & E-SO & -2.9 & 16.5 & -16.2 &     8.9 & C06 & 0.03 & 5.0 &  SDSS \\
       VCC1489 &    E & -5.0 & 16.5 & -15.2 &     8.6 & C06 & 0.05 & 5.6 &  SDSS \\
       VCC1528 &    E & -4.2 & 16.5 & -16.5 &     9.4 & C06 &    * & 5.7 &  SDSS \\
       VCC1539 &    E & -5.0 & 16.5 & -15.6 &     8.6 & C06 & 0.23 & 6.3 &  SDSS \\
       VCC1545 &    E & -3.6 & 16.5 & -16.3 &     8.9 & C06 & 0.05 & 5.9 &  SDSS \\
       NGC4540 & SABc &  6.1 & 19.9 & -19.0 & \nodata & B02 & 0.07 & 6.2 &  None \\
       VCC1619 &   S0 & -2.0 & 16.5 & -18.6 &    10.2 & C06 & 0.32 & 8.1 & HFS97 \\
       VCC1627 & E-SO & -3.3 & 16.5 & -15.6 &     9.2 & C06 & 0.20 & 7.4 &  SDSS \\
       VCC1630 &    E & -4.8 & 16.5 & -18.2 &    10.2 & C06 & 0.50 & 8.0 &  None \\
       VCC1661 &    E & -5.0 & 16.5 & -15.3 &     8.6 & C06 & 0.08 & 6.5 &  SDSS \\
       VCC1695 & S0-a & -0.8 & 16.5 & -16.4 &     9.3 & C06 &    * & 5.8 &  SDSS \\
       VCC1720 &   S0 & -2.0 & 16.5 & -18.7 &    10.3 & C06 & 0.09 & 7.6 & HFS97 \\
       VCC1826 &    E & -3.6 & 16.5 & -16.0 &     8.8 & C06 &    * & 6.7 &  SDSS \\
       VCC1828 &   Sc &  5.4 & 16.5 & -15.9 &     8.8 & C06 & 0.06 & 6.0 &  SDSS \\
       VCC1861 &    E & -4.8 & 16.5 & -16.6 &     9.3 & C06 & 0.14 & 6.6 &  SDSS \\
       VCC1871 &   S0 & -2.2 & 16.5 & -17.2 & \nodata & C06 & 0.12 & 7.3 &  SDSS \\
       NGC4618 &  SBm &  8.6 & 10.7 & -18.8 &     9.6 & B02 & 0.10 & 6.0 &  S\&H \\
       VCC1883 &   S0 & -2.0 & 16.5 & -19.0 &    10.4 & C06 &    * & 7.2 & HFS97 \\
       VCC1886 &    E & -5.0 & 16.5 & -16.1 & \nodata & C06 & 0.04 & 5.8 &  SDSS \\
       VCC1895 &    E & -3.5 & 16.5 & -16.1 &     9.0 & C06 &    * & 5.1 &  None \\
       NGC4625 & SABm &  8.8 & 11.7 & -17.3 &     9.4 & B02 & 0.10 & 5.6 &  SDSS \\
       VCC1910 &    E & -4.8 & 16.5 & -16.8 &     9.5 & C06 & 0.04 & 6.8 &  SDSS \\
       VCC1913 & S0-a & -1.5 & 16.5 & -17.8 &     9.9 & C06 & 0.60 & 7.9 &  None \\
       VCC2019 &    E & -3.5 & 16.5 & -16.4 &     9.2 & C06 & 0.04 & 6.6 &  None \\
       VCC2048 &    E & -4.7 & 16.5 & -17.2 &     9.6 & C06 & 0.04 & 6.1 &  None \\
       VCC2050 &    E & -5.0 & 16.5 & -15.8 & \nodata & C06 & 0.07 & 5.6 &  SDSS \\
       NGC4701 &   Sc &  5.8 & 11.1 & -17.4 &     9.3 & B02 & 0.04 & 6.5 &  None \\
       NGC4750 &  Sab &  2.4 & 27.2 & -20.3 &    10.6 & C02 & 0.14 & 8.1 & HFS97 \\
       NGC4775 &  Scd &  6.9 & 22.5 & -20.2 & \nodata & B02 & 0.06 & 7.6 &  None \\
       NGC4806 &   Sc &  4.9 & 32.7 & -19.0 & \nodata & C02 & 0.05 & 7.0 &  None \\
       NGC4980 &   Sa &  1.1 & 19.1 & -17.8 &     9.4 & C02 & 0.05 & 6.4 &  None \\
       NGC5023 &   Sc &  6.0 &  5.4 & -15.9 &     8.7 & S06 & 0.32 & 5.3 &  None \\
    ESO508-G34 & SABd &  8.3 & 26.2 & -17.2 &     8.8 & C98 & 0.06 & 6.4 &  None \\
       NGC5068 &   Sc &  6.0 &  8.7 & -19.0 &    10.0 & B02 & 0.11 & 6.2 &  None \\
       NGC5188 &   Sb &  3.0 & 32.7 & -19.6 &    10.7 & C02 & 0.20 & 8.1 &  None \\
       UGC8516 &   Sc &  5.9 & 16.7 & -17.3 &     9.3 & B02 & 0.09 & 5.8 &  None \\
       NGC5377 &   Sa &  1.1 & 31.0 & -20.3 &    10.9 & C02 & 0.18 & 8.6 & HFS97 \\
        IC4390 & SABb &  4.3 & 27.8 & -18.3 & \nodata & C02 & 0.17 & 8.1 &  None \\
       NGC5585 & SABc &  6.9 & 10.5 & -18.7 &     9.4 & B02 & 0.06 & 5.8 & HFS97 \\
       NGC5584 & SABc &  5.9 & 24.3 & -19.4 &    10.1 & B02 & 0.08 & 5.1 &  SDSS \\
       NGC5678 & SABb &  3.3 & 28.3 & -20.3 &    10.8 & C02 & 0.18 & 8.1 &  S\&H \\
       NGC5669 & SABc &  6.0 & 21.2 & -19.0 & \nodata & B02 & 0.05 & 6.5 &  S\&H \\
       NGC5774 & SABc &  6.9 & 23.7 & -18.8 &     9.8 & B02 & 0.12 & 5.5 &  SDSS \\
       NGC5789 &  SBd &  7.8 & 28.7 & -18.3 & \nodata & B02 & 0.07 & 5.5 &  None \\
       NGC5806 &   Sb &  3.1 & 21.4 & -19.3 &    10.5 & C02 & 0.21 & 8.1 &  S\&H \\
       NGC5879 &  Sbc &  3.5 & 15.0 & -18.8 &    10.0 & C97 & 0.17 & 7.2 &  S\&H \\
       NGC5964 & SBcd &  6.9 & 22.2 & -18.4 &     9.8 & B02 & 0.06 & 6.4 &  None \\
       NGC6000 & SBbc &  3.9 & 35.2 & -19.6 &    10.3 & C02 & 0.17 & 8.2 &  None \\
       NGC6239 &  SBb &  3.2 & 16.9 & -18.2 &     9.6 & C02 &    * & 6.7 &  SDSS \\
     ESO138-10 &   Sd &  7.8 & 13.5 & -18.9 &    10.3 & B02 & 0.09 & 7.0 &  None \\
       NGC6384 & SABb &  3.6 & 16.7 & -19.5 &    10.6 & C02 & 0.14 & 7.2 & HFS97 \\
       NGC6509 & SABc &  6.6 & 27.5 & -19.0 & \nodata & B02 & 0.04 & 6.4 &  None \\
       NGC6951 & SABb &  3.9 & 24.8 & -20.0 &    11.0 & C02 & 0.18 & 8.2 & HFS97 \\
        IC5052 & SBcd &  7.1 &  6.0 & -17.2 &     9.1 & S06 & 0.10 & 5.9 &  None \\
     ESO404-G3 & SBbc &  4.0 & 31.9 & -18.3 &     9.7 & C02 & 0.03 & 6.4 &  None \\
       NGC7162 &   Sc &  4.8 & 30.3 & -19.1 &    10.0 & C02 & 0.11 & 6.8 &  None \\
       NGC7188 & SBbc &  3.6 & 24.1 & -17.8 &     9.5 & C02 & 0.06 & 6.9 &  None \\
       NGC7259 &   Sb &  2.8 & 22.8 & -17.8 & \nodata & C02 & 0.06 & 6.9 &  None \\
        IC5256 &  SBd &  7.7 &  9.9 & -15.2 & \nodata & C98 &    * & 5.1 &  None \\
       NGC7418 &   Sc &  5.8 & 18.3 & -19.6 &    10.2 & B02 & 0.06 & 7.8 &  None \\
       NGC7421 &  Sbc &  3.8 & 23.7 & -19.2 &    10.2 & C02 & 0.11 & 6.9 &  None \\
       NGC7424 &   Sc &  5.9 & 10.8 & -19.1 &     9.6 & B02 & 0.10 & 6.1 &  None \\
        IC5271 &   Sb &  3.1 & 23.9 & -19.5 &    10.6 & C02 & 0.11 & 7.9 &  None \\
     ESO290-39 &  SBm &  8.9 & 19.1 & -16.2 & \nodata & B02 & 0.07 & 4.9 &  None \\
       NGC7513 &  SBb &  3.2 & 20.4 & -18.9 &    10.1 & C02 & 0.06 & 7.0 &  None \\
       NGC7690 &   Sb &  2.8 & 18.2 & -18.3 &     9.8 & C02 & 0.17 & 8.0 &  None \\
       NGC7689 & SABc &  6.0 & 24.8 & -19.8 &    10.3 & B02 & 0.08 & 6.9 &  None \\
    ESO240-G12 &   Sb &  3.1 & 23.0 & -17.3 &     9.3 & C02 & 0.03 & 6.0 &  None \\
      UGC12732 & SABm &  8.7 & 12.3 & -16.3 & \nodata & B02 & 0.07 & 5.8 &  None \\
      ESO241-6 &  SBm &  8.9 & 17.3 & -16.7 & \nodata & B02 & 0.06 & 5.3 &  None \\
       NGC7793 &  Scd &  7.4 &  3.9 & -18.3 &     9.6 & B02 & 0.10 & 6.9 &  None \\
\enddata
\tablenotetext{a}{Sources are ``B02''$=$\citet{boker02}, ``C06''$=$\citet{cote06}, ``C02''$=$\citet{carollo02}, ``C99''$=$\citet{carollo98}, ``C97''$=$\citet{carollo97}, ``S06''$=$\citet{seth06}.}
\tablenotetext{b}{NCs with sizes of `*' were unresolved by HST observations and thus have $r_{eff} \lesssim 0.03''$.}
\tablenotetext{c}{S\&H refers to spectra present in both SDSS and HFS97.}
\end{deluxetable}


\begin{thebibliography}{117}
\expandafter\ifx\csname natexlab\endcsname\relax\def\natexlab#1{#1}\fi

\bibitem[{{Adelman-McCarthy} {et~al.}(2007){Adelman-McCarthy}, {Ag{\"u}eros},
  {Allam}, {Anderson}, {Anderson}, {Annis}, {Bahcall}, {Bailer-Jones},
  {Baldry}, {Barentine}, {Beers}, {Belokurov}, {Berlind}, {Bernardi},
  {Blanton}, {Bochanski}, {Boroski}, {Bramich}, {Brewington}, {Brinchmann},
  {Brinkmann}, {Brunner}, {Budav{\'a}ri}, {Carey}, {Carliles}, {Carr},
  {Castander}, {Connolly}, {Cool}, {Cunha}, {Csabai}, {Dalcanton}, {Doi},
  {Eisenstein}, {Evans}, {Evans}, {Fan}, {Finkbeiner}, {Friedman}, {Frieman},
  {Fukugita}, {Gillespie}, {Gilmore}, {Glazebrook}, {Gray}, {Grebel}, {Gunn},
  {de Haas}, {Hall}, {Harvanek}, {Hawley}, {Hayes}, {Heckman}, {Hendry},
  {Hennessy}, {Hindsley}, {Hirata}, {Hogan}, {Hogg}, {Holtzman}, {Ichikawa},
  {Ichikawa}, {Ivezi{\'c}}, {Jester}, {Johnston}, {Jorgensen}, {Juri{\'c}},
  {Kauffmann}, {Kent}, {Kleinman}, {Knapp}, {Kniazev}, {Kron}, {Krzesinski},
  {Kuropatkin}, {Lamb}, {Lampeitl}, {Lee}, {Leger}, {Lima}, {Lin}, {Long},
  {Loveday}, {Lupton}, {Mandelbaum}, {Margon}, {Mart{\'{\i}}nez-Delgado},
  {Matsubara}, {McGehee}, {McKay}, {Meiksin}, {Munn}, {Nakajima}, {Nash},
  {Neilsen}, {Newberg}, {Nichol}, {Nieto-Santisteban}, {Nitta}, {Oyaizu},
  {Okamura}, {Ostriker}, {Padmanabhan}, {Park}, {Peoples}, {Pier}, {Pope},
  {Pourbaix}, {Quinn}, {Raddick}, {Re Fiorentin}, {Richards}, {Richmond},
  {Rix}, {Rockosi}, {Schlegel}, {Schneider}, {Scranton}, {Seljak}, {Sheldon},
  {Shimasaku}, {Silvestri}, {Smith}, {Smol{\v c}i{\'c}}, {Snedden}, {Stebbins},
  {Stoughton}, {Strauss}, {SubbaRao}, {Suto}, {Szalay}, {Szapudi}, {Szkody},
  {Tegmark}, {Thakar}, {Tremonti}, {Tucker}, {Uomoto}, {Vanden Berk},
  {Vandenberg}, {Vidrih}, {Vogeley}, {Voges}, {Vogt}, {Weinberg}, {West},
  {White}, {Wilhite}, {Yanny}, {Yocum}, {York}, {Zehavi}, {Zibetti}, \&
  {Zucker}}]{sloandr6}
{Adelman-McCarthy}, J.~K. {et~al.} 2007, astro-ph/0707.3413

\bibitem[{{Baldwin} {et~al.}(1981){Baldwin}, {Phillips}, \&
  {Terlevich}}]{baldwin81}
{Baldwin}, J.~A., {Phillips}, M.~M., \& {Terlevich}, R. 1981, \pasp, 93, 5

\bibitem[{{Baumgardt} {et~al.}(2003){Baumgardt}, {Makino}, {Hut}, {McMillan},
  \& {Portegies Zwart}}]{baumgardt03}
{Baumgardt}, H., {Makino}, J., {Hut}, P., {McMillan}, S., \& {Portegies Zwart},
  S. 2003, \apjl, 589, L25

\bibitem[{{Becker} {et~al.}(1995)}]{first}
{Becker}, R.~H., {et~al.} 1995, \apj, 450, 559

\bibitem[{{Becklin} \& {Neugebauer}(1968)}]{becklin68}
{Becklin}, E.~E., \& {Neugebauer}, G. 1968, \apj, 151, 145

\bibitem[{{Bedin} {et~al.}(2004){Bedin}, {Piotto}, {Anderson}, {Cassisi},
  {King}, {Momany}, \& {Carraro}}]{bedin04}
{Bedin}, L.~R., {Piotto}, G., {Anderson}, J., {Cassisi}, S., {King}, I.~R.,
  {Momany}, Y., \& {Carraro}, G. 2004, \apjl, 605, L125

\bibitem[{{Bekki} {et~al.}(2006){Bekki}, {Couch}, \& {Shioya}}]{bekki06}
{Bekki}, K., {Couch}, W.~J., \& {Shioya}, Y. 2006, \apjl, 642, L133

\bibitem[{{Bell} {et~al.}(2003){Bell}, {McIntosh}, {Katz}, \&
  {Weinberg}}]{bell03}
{Bell}, E.~F., {McIntosh}, D.~H., {Katz}, N., \& {Weinberg}, M.~D. 2003, \apjs,
  149, 289

\bibitem[{{Best}(2004)}]{best04}
{Best}, P.~N. 2004, \mnras, 351, 70

\bibitem[{{B{\"o}ker} {et~al.}(2002){B{\"o}ker}, {Laine}, {van der Marel},
  {Sarzi}, {Rix}, {Ho}, \& {Shields}}]{boker02}
{B{\"o}ker}, T., {Laine}, S., {van der Marel}, R.~P., {Sarzi}, M., {Rix},
  H.-W., {Ho}, L.~C., \& {Shields}, J.~C. 2002, \aj, 123, 1389

\bibitem[{{B{\"o}ker} {et~al.}(2004){B{\"o}ker}, {Sarzi}, {McLaughlin}, {van
  der Marel}, {Rix}, {Ho}, \& {Shields}}]{boker04a}
{B{\"o}ker}, T., {Sarzi}, M., {McLaughlin}, D.~E., {van der Marel}, R.~P.,
  {Rix}, H.-W., {Ho}, L.~C., \& {Shields}, J.~C. 2004, \aj, 127, 105

\bibitem[{{Bruzual} \& {Charlot}(2003)}]{bruzual03}
{Bruzual}, G., \& {Charlot}, S. 2003, \mnras, 344, 1000

\bibitem[{{Carollo} {et~al.}(1997){Carollo}, {Stiavelli}, {de Zeeuw}, \&
  {Mack}}]{carollo97}
{Carollo}, C.~M., {Stiavelli}, M., {de Zeeuw}, P.~T., \& {Mack}, J. 1997, \aj,
  114, 2366

\bibitem[{{Carollo} {et~al.}(1998){Carollo}, {Stiavelli}, \&
  {Mack}}]{carollo98}
{Carollo}, C.~M., {Stiavelli}, M., \& {Mack}, J. 1998, \aj, 116, 68

\bibitem[{{Carollo} {et~al.}(2002){Carollo}, {Stiavelli}, {Seigar}, {de Zeeuw},
  \& {Dejonghe}}]{carollo02}
{Carollo}, C.~M., {Stiavelli}, M., {Seigar}, M., {de Zeeuw}, P.~T., \&
  {Dejonghe}, H. 2002, \aj, 123, 159

\bibitem[{{Colina} {et~al.}(2002){Colina}, {Gonzalez Delgado}, {Mas-Hesse}, \&
  {Leitherer}}]{colina02}
{Colina}, L., {Gonzalez Delgado}, R., {Mas-Hesse}, J.~M., \& {Leitherer}, C.
  2002, \apj, 579, 545

\bibitem[{{Condon} {et~al.}(1998)}]{nvss}
{Condon}, J.~J., {et~al.} 1998, \aj, 115, 1693

\bibitem[{{C{\^o}t{\'e}} {et~al.}(2006){C{\^o}t{\'e}}, {Piatek}, {Ferrarese},
  {Jord{\'a}n}, {Merritt}, {Peng}, {Ha{\c s}egan}, {Blakeslee}, {Mei}, {West},
  {Milosavljevi{\'c}}, \& {Tonry}}]{cote06}
{C{\^o}t{\'e}}, P. {et~al.} 2006, \apjs, 165, 57

\bibitem[{Croft {et~al.}(2007)Croft, de~Vries, \& Becker}]{croft07}
Croft, S., de~Vries, W., \& Becker, R.~H. 2007, astro-ph/0708.0585

\bibitem[{{De Rijcke} {et~al.}(2006){De Rijcke}, {Prugniel}, {Simien}, \&
  {Dejonghe}}]{derijcke06}
{De Rijcke}, S., {Prugniel}, P., {Simien}, F., \& {Dejonghe}, H. 2006, \mnras,
  369, 1321

\bibitem[{{Decarli} {et~al.}(2007){Decarli}, {Gavazzi}, {Arosio}, {Cortese},
  {Boselli}, {Bonfanti}, \& {Colpi}}]{decarli07}
{Decarli}, R., {Gavazzi}, G., {Arosio}, I., {Cortese}, L., {Boselli}, A.,
  {Bonfanti}, C., \& {Colpi}, M. 2007, astro-ph/0707.0999, 707

\bibitem[{{Dubus} {et~al.}(2004){Dubus}, {Charles}, \& {Long}}]{dubus04}
{Dubus}, G., {Charles}, P.~A., \& {Long}, K.~S. 2004, \aap, 425, 95

\bibitem[{{Dudik} {et~al.}(2005){Dudik}, {Satyapal}, {Gliozzi}, \&
  {Sambruna}}]{dudik05}
{Dudik}, R.~P., {Satyapal}, S., {Gliozzi}, M., \& {Sambruna}, R.~M. 2005, \apj,
  620, 113

\bibitem[{{Efron}(1982)}]{dasboot}
{Efron}, B. 1982, {The Jackknife, the Bootstrap and other resampling plans}
  (CBMS-NSF Regional Conference Series in Applied Mathematics, Philadelphia:
  Society for Industrial and Applied Mathematics (SIAM), 1982)

\bibitem[{{Elvis} {et~al.}(1994){Elvis}, {Wilkes}, {McDowell}, {Green},
  {Bechtold}, {Willner}, {Oey}, {Polomski}, \& {Cutri}}]{elvis94}
{Elvis}, M. {et~al.} 1994, \apjs, 95, 1

\bibitem[{{Fabbiano}(2006)}]{fabbiano06}
{Fabbiano}, G. 2006, \araa, 44, 323

\bibitem[{{Ferrarese} {et~al.}(2006){Ferrarese}, {C{\^o}t{\'e}}, {Dalla
  Bont{\`a}}, {Peng}, {Merritt}, {Jord{\'a}n}, {Blakeslee}, {Ha{\c s}egan},
  {Mei}, {Piatek}, {Tonry}, \& {West}}]{ferrarese06}
{Ferrarese}, L. {et~al.} 2006, \apjl, 644, L21

\bibitem[{{Filippenko} \& {Ho}(2003)}]{filippenko03}
{Filippenko}, A.~V., \& {Ho}, L.~C. 2003, \apjl, 588, L13

\bibitem[{{Filippenko} \& {Sargent}(1985)}]{filippenko85}
{Filippenko}, A.~V., \& {Sargent}, W.~L.~W. 1985, \apjs, 57, 503

\bibitem[{{Filippenko} \& {Sargent}(1989)}]{filippenko89}
---. 1989, \apjl, 342, L11

\bibitem[{{Flesch} \& {Hardcastle}(2004)}]{flesch04}
{Flesch}, E., \& {Hardcastle}, M.~J. 2004, \aap, 427, 387

\bibitem[{{Gebhardt} {et~al.}(2001){Gebhardt}, {Lauer}, {Kormendy}, {Pinkney},
  {Bower}, {Green}, {Gull}, {Hutchings}, {Kaiser}, {Nelson}, {Richstone}, \&
  {Weistrop}}]{gebhardt01}
{Gebhardt}, K. {et~al.} 2001, \aj, 122, 2469

\bibitem[{{Gebhardt} {et~al.}(2002){Gebhardt}, {Rich}, \& {Ho}}]{gebhardt02}
{Gebhardt}, K., {Rich}, R.~M., \& {Ho}, L.~C. 2002, \apjl, 578, L41

\bibitem[{{Gebhardt} {et~al.}(2005){Gebhardt}, {Rich}, \& {Ho}}]{gebhardt05}
---. 2005, \apj, 634, 1093

\bibitem[{{Genzel} {et~al.}(1996){Genzel}, {Thatte}, {Krabbe}, {Kroker}, \&
  {Tacconi-Garman}}]{genzel96}
{Genzel}, R., {Thatte}, N., {Krabbe}, A., {Kroker}, H., \& {Tacconi-Garman},
  L.~E. 1996, \apj, 472, 153

\bibitem[{{Ghez} {et~al.}(2005){Ghez}, {Salim}, {Hornstein}, {Tanner}, {Lu},
  {Morris}, {Becklin}, \& {Duch{\^e}ne}}]{ghez05}
{Ghez}, A.~M., {Salim}, S., {Hornstein}, S.~D., {Tanner}, A., {Lu}, J.~R.,
  {Morris}, M., {Becklin}, E.~E., \& {Duch{\^e}ne}, G. 2005, \apj, 620, 744

\bibitem[{{Ghosh} {et~al.}(2006){Ghosh}, {Suleymanov}, {Bikmaev}, {Shimansky},
  \& {Sakhibullin}}]{ghosh06}
{Ghosh}, K.~K., {Suleymanov}, V., {Bikmaev}, I., {Shimansky}, S., \&
  {Sakhibullin}, N. 2006, \mnras, 371, 1587

\bibitem[{{Gonzalez Delgado} {et~al.}(2007){Gonzalez Delgado}, {Perez}, {Cid
  Fernandes}, \& {Schmitt}}]{gonzalezdelgado07}
{Gonzalez Delgado}, R.~M., {Perez}, E., {Cid Fernandes}, R., \& {Schmitt}, H.
  2007, ArXiv e-prints, 710

\bibitem[{{Gonz{\'a}lez-Mart{\'{\i}}n}
  {et~al.}(2006){Gonz{\'a}lez-Mart{\'{\i}}n}, {Masegosa}, {M{\'a}rquez},
  {Guerrero}, \& {Dultzin-Hacyan}}]{gonzalezmartin06}
{Gonz{\'a}lez-Mart{\'{\i}}n}, O., {Masegosa}, J., {M{\'a}rquez}, I.,
  {Guerrero}, M.~A., \& {Dultzin-Hacyan}, D. 2006, \aap, 460, 45

\bibitem[{{Graham} \& {Driver}(2007)}]{graham07}
{Graham}, A.~W., \& {Driver}, S.~P. 2007, \apj, 655, 77

\bibitem[{{Graham} \& {Guzm{\'a}n}(2003)}]{graham03}
{Graham}, A.~W., \& {Guzm{\'a}n}, R. 2003, \aj, 125, 2936

\bibitem[{{Greene} \& {Ho}(2007)}]{greene07}
{Greene}, J.~E., \& {Ho}, L.~C. 2007, \apj, 667, 131

\bibitem[{{Haberl} {et~al.}(2000)}]{haberl00}
{Haberl}, F., {et~al.} 2000, \aaps, 142, 41

\bibitem[{{Ho}(1999)}]{ho99}
{Ho}, L.~C. 1999, \apj, 516, 672

\bibitem[{{Ho} {et~al.}(2001){Ho}, {Feigelson}, {Townsley}, {Sambruna},
  {Garmire}, {Brandt}, {Filippenko}, {Griffiths}, {Ptak}, \& {Sargent}}]{ho01}
{Ho}, L.~C. {et~al.} 2001, \apjl, 549, L51

\bibitem[{{Ho} {et~al.}(1997{\natexlab{a}}){Ho}, {Filippenko}, \&
  {Sargent}}]{ho97a}
{Ho}, L.~C., {Filippenko}, A.~V., \& {Sargent}, W.~L.~W. 1997{\natexlab{a}},
  \apjs, 112, 315

\bibitem[{{Ho} {et~al.}(1997{\natexlab{b}}){Ho}, {Filippenko}, \&
  {Sargent}}]{ho97b}
---. 1997{\natexlab{b}}, \apj, 487, 568

\bibitem[{{Ho} {et~al.}(2003){Ho}, {Terashima}, \& {Ulvestad}}]{ho03a}
{Ho}, L.~C., {Terashima}, Y., \& {Ulvestad}, J.~S. 2003, \apj, 589, 783

\bibitem[{{Ho}(2004)}]{ho04}
{Ho}, L.~C.~W. 2004, in Coevolution of Black Holes and Galaxies, ed. L.~C.
  {Ho}, 292--+

\bibitem[{{Jensen} {et~al.}(2003){Jensen}, {Tonry}, {Barris}, {Thompson},
  {Liu}, {Rieke}, {Ajhar}, \& {Blakeslee}}]{jensen03}
{Jensen}, J.~B., {Tonry}, J.~L., {Barris}, B.~J., {Thompson}, R.~I., {Liu},
  M.~C., {Rieke}, M.~J., {Ajhar}, E.~A., \& {Blakeslee}, J.~P. 2003, \apj, 583,
  712

\bibitem[{{Jones} {et~al.}(2004){Jones}, {Saunders}, {Colless}, {Read},
  {Parker}, {Watson}, {Campbell}, {Burkey}, {Mauch}, {Moore}, {Hartley},
  {Cass}, {James}, {Russell}, {Fiegert}, {Dawe}, {Huchra}, {Jarrett}, {Lahav},
  {Lucey}, {Mamon}, {Proust}, {Sadler}, \& {Wakamatsu}}]{jones04}
{Jones}, D.~H. {et~al.} 2004, \mnras, 355, 747

\bibitem[{{Kauffmann} {et~al.}(2003){Kauffmann}, {Heckman}, {Tremonti},
  {Brinchmann}, {Charlot}, {White}, {Ridgway}, {Brinkmann}, {Fukugita}, {Hall},
  {Ivezi{\'c}}, {Richards}, \& {Schneider}}]{kauffmann03}
{Kauffmann}, G. {et~al.} 2003, \mnras, 346, 1055

\bibitem[{{Kennicutt}(1998)}]{kennicutt98a}
{Kennicutt}, Jr., R.~C. 1998, \apj, 498, 541

\bibitem[{{Kewley} {et~al.}(2001){Kewley}, {Dopita}, {Sutherland}, {Heisler},
  \& {Trevena}}]{kewley01}
{Kewley}, L.~J., {Dopita}, M.~A., {Sutherland}, R.~S., {Heisler}, C.~A., \&
  {Trevena}, J. 2001, \apj, 556, 121

\bibitem[{{Kewley} {et~al.}(2006){Kewley}, {Groves}, {Kauffmann}, \&
  {Heckman}}]{kewley06}
{Kewley}, L.~J., {Groves}, B., {Kauffmann}, G., \& {Heckman}, T. 2006, \mnras,
  372, 961

\bibitem[{{Kormendy} \& {McClure}(1993)}]{kormendy93}
{Kormendy}, J., \& {McClure}, R.~D. 1993, \aj, 105, 1793

\bibitem[{{Li} {et~al.}(2007){Li}, {Haiman}, \& {Mac Low}}]{li07}
{Li}, Y., {Haiman}, Z., \& {Mac Low}, M.-M. 2007, \apj, 663, 61

\bibitem[{{Lotz} {et~al.}(2001){Lotz}, {Telford}, {Ferguson}, {Miller},
  {Stiavelli}, \& {Mack}}]{lotz01}
{Lotz}, J.~M., {Telford}, R., {Ferguson}, H.~C., {Miller}, B.~W., {Stiavelli},
  M., \& {Mack}, J. 2001, \apj, 552, 572

\bibitem[{{Maccarone} {et~al.}(2007){Maccarone}, {Kundu}, {Zepf}, \&
  {Rhode}}]{maccarone07}
{Maccarone}, T.~J., {Kundu}, A., {Zepf}, S.~E., \& {Rhode}, K.~L. 2007, \nat,
  445, 183

\bibitem[{{Maoz} {et~al.}(2005){Maoz}, {Nagar}, {Falcke}, \& {Wilson}}]{maoz05}
{Maoz}, D., {Nagar}, N.~M., {Falcke}, H., \& {Wilson}, A.~S. 2005, \apj, 625,
  699

\bibitem[{{Marconi} \& {Hunt}(2003)}]{marconi03}
{Marconi}, A., \& {Hunt}, L.~K. 2003, \apjl, 589, L21

\bibitem[{{Matthews} {et~al.}(1999){Matthews}, {Gallagher}, {Krist}, {Watson},
  {Burrows}, {Griffiths}, {Hester}, {Trauger}, {Ballester}, {Clarke}, {Crisp},
  {Evans}, {Hoessel}, {Holtzman}, {Mould}, {Scowen}, {Stapelfeldt}, \&
  {Westphal}}]{matthews99}
{Matthews}, L.~D. {et~al.} 1999, \aj, 118, 208

\bibitem[{{McLaughlin} {et~al.}(2006{\natexlab{a}}){McLaughlin}, {Anderson},
  {Meylan}, {Gebhardt}, {Pryor}, {Minniti}, \& {Phinney}}]{mclaughlin06a}
{McLaughlin}, D.~E., {Anderson}, J., {Meylan}, G., {Gebhardt}, K., {Pryor}, C.,
  {Minniti}, D., \& {Phinney}, S. 2006{\natexlab{a}}, \apjs, 166, 249

\bibitem[{{McLaughlin} {et~al.}(2006{\natexlab{b}}){McLaughlin}, {King}, \&
  {Nayakshin}}]{mclaughlin06b}
{McLaughlin}, D.~E., {King}, A.~R., \& {Nayakshin}, S. 2006{\natexlab{b}},
  \apjl, 650, L37

\bibitem[{{Mei} {et~al.}(2007){Mei}, {Blakeslee}, {C{\^o}t{\'e}}, {Tonry},
  {West}, {Ferrarese}, {Jord{\'a}n}, {Peng}, {Anthony}, \& {Merritt}}]{mei07}
{Mei}, S. {et~al.} 2007, \apj, 655, 144

\bibitem[{{Meylan} {et~al.}(2001){Meylan}, {Sarajedini}, {Jablonka},
  {Djorgovski}, {Bridges}, \& {Rich}}]{meylan01}
{Meylan}, G., {Sarajedini}, A., {Jablonka}, P., {Djorgovski}, S.~G., {Bridges},
  T., \& {Rich}, R.~M. 2001, \aj, 122, 830

\bibitem[{{Mickaelian} {et~al.}(2006)}]{mick06}
{Mickaelian}, A.~M., {et~al.} 2006, \aap, 449, 425

\bibitem[{{Mihos} \& {Hernquist}(1994)}]{mihos94}
{Mihos}, J.~C., \& {Hernquist}, L. 1994, \apjl, 437, L47

\bibitem[{{Miller} \& {Colbert}(2004)}]{miller04}
{Miller}, M.~C., \& {Colbert}, E.~J.~M. 2004, International Journal of Modern
  Physics D, 13, 1

\bibitem[{{Milosavljevi{\'c}}(2004)}]{milosavljevic04}
{Milosavljevi{\'c}}, M. 2004, \apjl, 605, L13

\bibitem[{{Milosavljevi{\'c}} \& {Merritt}(2001)}]{milosavljevic01}
{Milosavljevi{\'c}}, M., \& {Merritt}, D. 2001, \apj, 563, 34

\bibitem[{{Moran} {et~al.}(1996){Moran}, {Halpern}, \& {Helfand}}]{moran96}
{Moran}, E.~C., {Halpern}, J.~P., \& {Helfand}, D.~J. 1996, \apjs, 106, 341

\bibitem[{{Mukai}(1993)}]{mukai93}
{Mukai}, K. 1993, Legacy, vol.~3, p.21-31, 3, 21

\bibitem[{{Nagar} {et~al.}(2005){Nagar}, {Falcke}, \& {Wilson}}]{nagar05}
{Nagar}, N.~M., {Falcke}, H., \& {Wilson}, A.~S. 2005, \aap, 435, 521

\bibitem[{{Narayan} {et~al.}(1998){Narayan}, {Mahadevan}, {Grindlay}, {Popham},
  \& {Gammie}}]{narayan98}
{Narayan}, R., {Mahadevan}, R., {Grindlay}, J.~E., {Popham}, R.~G., \&
  {Gammie}, C. 1998, \apj, 492, 554

\bibitem[{{Noyola} {et~al.}(2006){Noyola}, {Gebhardt}, \&
  {Bergmann}}]{noyola06}
{Noyola}, E., {Gebhardt}, K., \& {Bergmann}, M. 2006, in Astronomical Society
  of the Pacific Conference Series, Vol. 352, New Horizons in Astronomy: Frank
  N. Bash Symposium, ed. S.~J. {Kannappan}, S.~{Redfield}, J.~E.
  {Kessler-Silacci}, M.~{Landriau}, \& N.~{Drory}, 269--+

\bibitem[{{Osterbrock}(1989)}]{osterbrock89}
{Osterbrock}, D.~E. 1989, {Astrophysics of gaseous nebulae and active galactic
  nuclei} (University Science Books)

\bibitem[{{Panessa} {et~al.}(2006){Panessa}, {Bassani}, {Cappi}, {Dadina},
  {Barcons}, {Carrera}, {Ho}, \& {Iwasawa}}]{panessa06}
{Panessa}, F., {Bassani}, L., {Cappi}, M., {Dadina}, M., {Barcons}, X.,
  {Carrera}, F.~J., {Ho}, L.~C., \& {Iwasawa}, K. 2006, \aap, 455, 173

\bibitem[{{Panzera} {et~al.}(2003)}]{panzera03}
{Panzera}, M.~R., {et~al.} 2003, \aap, 399, 351

\bibitem[{{Paturel} {et~al.}(2003){Paturel}, {Theureau}, {Bottinelli},
  {Gouguenheim}, {Coudreau-Durand}, {Hallet}, \& {Petit}}]{paturel03}
{Paturel}, G., {Theureau}, G., {Bottinelli}, L., {Gouguenheim}, L.,
  {Coudreau-Durand}, N., {Hallet}, N., \& {Petit}, C. 2003, \aap, 412, 57

\bibitem[{{Peterson} {et~al.}(2005){Peterson}, {Bentz}, {Desroches},
  {Filippenko}, {Ho}, {Kaspi}, {Laor}, {Maoz}, {Moran}, {Pogge}, \&
  {Quillen}}]{peterson05}
{Peterson}, B.~M. {et~al.} 2005, \apj, 632, 799

\bibitem[{{Pooley} \& {Rappaport}(2006)}]{pooley06}
{Pooley}, D., \& {Rappaport}, S. 2006, \apjl, 644, L45

\bibitem[{{Portegies Zwart} {et~al.}(2004){Portegies Zwart}, {Baumgardt},
  {Hut}, {Makino}, \& {McMillan}}]{portegieszwart04}
{Portegies Zwart}, S.~F., {Baumgardt}, H., {Hut}, P., {Makino}, J., \&
  {McMillan}, S.~L.~W. 2004, \nat, 428, 724

\bibitem[{{Ptak} \& {Griffiths}(2003)}]{ptak03}
{Ptak}, A., \& {Griffiths}, R. 2003, in Astronomical Society of the Pacific
  Conference Series, Vol. 295, Astronomical Data Analysis Software and Systems
  XII, ed. H.~E. {Payne}, R.~I. {Jedrzejewski}, \& R.~N. {Hook}, 465--+

\bibitem[{{Rasio} {et~al.}(2006){Rasio}, {Baumgardt}, {Corongiu}, {D'Antona},
  {Fabbiano}, {Fregeau}, {Gebhardt}, {Heinke}, {Hut}, {Ivanova}, {Maccarone},
  {Ransom}, \& {Webb}}]{rasio06}
{Rasio}, F.~A. {et~al.} 2006, ArXiv Astrophysics e-prints

\bibitem[{{Ravindranath} {et~al.}(2001){Ravindranath}, {Ho}, {Peng},
  {Filippenko}, \& {Sargent}}]{ravindranath01}
{Ravindranath}, S., {Ho}, L.~C., {Peng}, C.~Y., {Filippenko}, A.~V., \&
  {Sargent}, W.~L.~W. 2001, \aj, 122, 653

\bibitem[{{ROSAT Scientific Team}(2000)}]{rosat00}
{ROSAT Scientific Team}. 2000, VizieR Online Data Catalog, 9028, 0

\bibitem[{{Rossa} {et~al.}(2006){Rossa}, {van der Marel}, {B{\"o}ker},
  {Gerssen}, {Ho}, {Rix}, {Shields}, \& {Walcher}}]{rossa06}
{Rossa}, J., {van der Marel}, R.~P., {B{\"o}ker}, T., {Gerssen}, J., {Ho},
  L.~C., {Rix}, H.-W., {Shields}, J.~C., \& {Walcher}, C.-J. 2006, \aj, 132,
  1074

\bibitem[{{Satyapal} {et~al.}(2007){Satyapal}, {Vega}, {Heckman}, {O'Halloran},
  \& {Dudik}}]{satyapal07}
{Satyapal}, S., {Vega}, D., {Heckman}, T., {O'Halloran}, B., \& {Dudik}, R.
  2007, \apjl, 663, L9

\bibitem[{{Scarlata} {et~al.}(2004){Scarlata}, {Stiavelli}, {Hughes}, {Axon},
  {Alonso-Herrero}, {Atkinson}, {Batcheldor}, {Binney}, {Capetti}, {Carollo},
  {Dressel}, {Gerssen}, {Macchetto}, {Maciejewski}, {Marconi}, {Merrifield},
  {Ruiz}, {Sparks}, {Tsvetanov}, \& {van der Marel}}]{scarlata04}
{Scarlata}, C. {et~al.} 2004, \aj, 128, 1124

\bibitem[{{Sch{\"o}del} {et~al.}(2007){Sch{\"o}del}, {Eckart}, {Alexander},
  {Merritt}, {Genzel}, {Sternberg}, {Meyer}, {Kul}, {Moultaka}, {Ott}, \&
  {Straubmeier}}]{schodel07}
{Sch{\"o}del}, R. {et~al.} 2007, \aap, 469, 125

\bibitem[{{Seth} {et~al.}(2006){Seth}, {Dalcanton}, {Hodge}, \&
  {Debattista}}]{seth06}
{Seth}, A.~C., {Dalcanton}, J.~J., {Hodge}, P.~W., \& {Debattista}, V.~P. 2006,
  \aj, 132, 2539

\bibitem[{{Shapley} {et~al.}(2001){Shapley}, {Fabbiano}, \&
  {Eskridge}}]{shapley01}
{Shapley}, A., {Fabbiano}, G., \& {Eskridge}, P.~B. 2001, \apjs, 137, 139

\bibitem[{{Shields} {et~al.}(2008){Shields}, {Walcher}, {Boeker}, {Ho}, {Rix},
  \& {van der Marel}}]{shields08}
{Shields}, J., {Walcher}, C.~J.~M., {Boeker}, T., {Ho}, L.~C., {Rix}, H.-W., \&
  {van der Marel}, R.~P. 2008, {\it submitted}

\bibitem[{{Shields} {et~al.}(2007){Shields}, {Rix}, {Sarzi}, {Barth},
  {Filippenko}, {Ho}, {McIntosh}, {Rudnick}, \& {Sargent}}]{shields07}
{Shields}, J.~C. {et~al.} 2007, \apj, 654, 125

\bibitem[{{Sivakoff} {et~al.}(2007){Sivakoff}, {Jord{\'a}n}, {Sarazin},
  {Blakeslee}, {C{\^o}t{\'e}}, {Ferrarese}, {Juett}, {Mei}, \&
  {Peng}}]{sivakoff07}
{Sivakoff}, G.~R. {et~al.} 2007, \apj, 660, 1246

\bibitem[{{Terashima} \& {Wilson}(2003)}]{terashima03}
{Terashima}, Y., \& {Wilson}, A.~S. 2003, \apj, 583, 145

\bibitem[{{Thim} {et~al.}(2004){Thim}, {Hoessel}, {Saha}, {Claver}, {Dolphin},
  \& {Tammann}}]{thim04}
{Thim}, F., {Hoessel}, J.~G., {Saha}, A., {Claver}, J., {Dolphin}, A., \&
  {Tammann}, G.~A. 2004, \aj, 127, 2322

\bibitem[{{Tonry} {et~al.}(2001){Tonry}, {Dressler}, {Blakeslee}, {Ajhar},
  {Fletcher}, {Luppino}, {Metzger}, \& {Moore}}]{tonry01}
{Tonry}, J.~L., {Dressler}, A., {Blakeslee}, J.~P., {Ajhar}, E.~A., {Fletcher},
  A.~B., {Luppino}, G.~A., {Metzger}, M.~R., \& {Moore}, C.~B. 2001, \apj, 546,
  681

\bibitem[{{Tremaine} {et~al.}(2002){Tremaine}, {Gebhardt}, {Bender}, {Bower},
  {Dressler}, {Faber}, {Filippenko}, {Green}, {Grillmair}, {Ho}, {Kormendy},
  {Lauer}, {Magorrian}, {Pinkney}, \& {Richstone}}]{tremaine02}
{Tremaine}, S. {et~al.} 2002, \apj, 574, 740

\bibitem[{{Tremaine} {et~al.}(1975){Tremaine}, {Ostriker}, \&
  {Spitzer}}]{tremaine75}
{Tremaine}, S.~D., {Ostriker}, J.~P., \& {Spitzer}, L. 1975, \apj, 196, 407

\bibitem[{{Tremonti} {et~al.}(2004){Tremonti}, {Heckman}, {Kauffmann},
  {Brinchmann}, {Charlot}, {White}, {Seibert}, {Peng}, {Schlegel}, {Uomoto},
  {Fukugita}, \& {Brinkmann}}]{tremonti04}
{Tremonti}, C.~A. {et~al.} 2004, \apj, 613, 898

\bibitem[{{Trenti}(2006)}]{trenti06}
{Trenti}, M. 2006, astro-ph/0612040

\bibitem[{{Ulvestad} {et~al.}(2007){Ulvestad}, {Greene}, \& {Ho}}]{ulvestad07}
{Ulvestad}, J.~S., {Greene}, J.~E., \& {Ho}, L.~C. 2007, \apjl, 661, L151

\bibitem[{{Valluri} {et~al.}(2005){Valluri}, {Ferrarese}, {Merritt}, \&
  {Joseph}}]{valluri05}
{Valluri}, M., {Ferrarese}, L., {Merritt}, D., \& {Joseph}, C.~L. 2005, \apj,
  628, 137

\bibitem[{{van den Bosch} {et~al.}(2006){van den Bosch}, {de Zeeuw},
  {Gebhardt}, {Noyola}, \& {van de Ven}}]{vandenbosch06}
{van den Bosch}, R., {de Zeeuw}, T., {Gebhardt}, K., {Noyola}, E., \& {van de
  Ven}, G. 2006, \apj, 641, 852

\bibitem[{{van der Marel}(2004)}]{vandermarel04}
{van der Marel}, R.~P. 2004, in Coevolution of Black Holes and Galaxies, ed.
  L.~C. {Ho}, 37--+

\bibitem[{{V{\'e}ron-Cetty} {et~al.}(2004){V{\'e}ron-Cetty}, {Balayan},
  {Mickaelian}, {Mujica}, {Chavushyan}, {Hakopian}, {Engels}, {V{\'e}ron},
  {Zickgraf}, {Voges}, \& {Xu}}]{veron04}
{V{\'e}ron-Cetty}, M.-P. {et~al.} 2004, \aap, 414, 487

\bibitem[{{Voges} {et~al.}(1999)}]{voges99}
{Voges}, W., {et~al.} 1999, \aap, 349, 389

\bibitem[{{Voges} {et~al.}(2000)}]{fsc}
---. 2000, VizieR Online Data Catalog, 9029, 0

\bibitem[{{Walcher} {et~al.}(2006){Walcher}, {B{\"o}ker}, {Charlot}, {Ho},
  {Rix}, {Rossa}, {Shields}, \& {van der Marel}}]{walcher06}
{Walcher}, C.~J., {B{\"o}ker}, T., {Charlot}, S., {Ho}, L.~C., {Rix}, H.-W.,
  {Rossa}, J., {Shields}, J.~C., \& {van der Marel}, R.~P. 2006, \apj, 649, 692

\bibitem[{{Walcher} {et~al.}(2005){Walcher}, {van der Marel}, {McLaughlin},
  {Rix}, {B{\"o}ker}, {H{\"a}ring}, {Ho}, {Sarzi}, \& {Shields}}]{walcher05}
{Walcher}, C.~J. {et~al.} 2005, \apj, 618, 237

\bibitem[{{Wehner} \& {Harris}(2006)}]{wehner06}
{Wehner}, E.~H., \& {Harris}, W.~E. 2006, \apjl, 644, L17

\bibitem[{{White} {et~al.}(1996)}]{white96}
{White}, N.~E., {et~al.} 1996, VizieR Online Data Catalog, 9012, 0

\bibitem[{{White} {et~al.}(2007)}]{white07}
{White}, R.~L., {et~al.} 2007, \apj, 654, 99

\bibitem[{{York} {et~al.}(2000)}]{york00}
{York}, D.~G., {et~al.} 2000, \aj, 120, 1579

\bibitem[{{Yukita} {et~al.}(2007){Yukita}, {Swartz}, {Soria}, \&
  {Tennant}}]{yukita07}
{Yukita}, M., {Swartz}, D.~A., {Soria}, R., \& {Tennant}, A.~F. 2007, \apj,
  664, 277

\end{thebibliography}
\end{document}